\documentclass{article}
\usepackage{titlesec}

\titleformat{\paragraph}
{\normalfont\normalsize\bfseries}{\theparagraph}{1em}{}
\titlespacing*{\paragraph}
{0pt}{3.25ex plus 1ex minus .2ex}{1.5ex plus .2ex}
\usepackage{amsfonts}
\usepackage{amsmath}
\usepackage{amssymb}
\usepackage{graphicx}
\usepackage{epstopdf}
\usepackage{multirow}
\usepackage{hhline}
\usepackage{bm}
\usepackage{authblk}
\usepackage{titling}

\newcommand{\ket}[1]{|#1\rangle}

\input{tcilatex}

\numberwithin{equation}{section}

\begin{document}

\title{Introduction to Quantum Electromagnetic Circuits}

\author{Uri Vool\thanks{e-mail: uri.vool@yale.edu} }

\author{Michel Devoret\thanks{e-mail: michel.devoret@yale.edu \\ \hspace*{1.6em}Address: 15 Prospect St., New Haven, CT 06511 \\ \hspace*{1.6em}Phone: 203-432-2210 \\ \hspace*{1.6em}Fax: 203-432-4283} }

\affil{Department of Applied Physics, \\Yale University, New Haven, CT 06520} 
\date{}
\begin{titlingpage}
\maketitle
\abstract{The article is a short opinionated review of the quantum treatment of electromagnetic circuits, with no pretension to exhaustiveness. This review, which is an updated and modernized version of a previous set of Les Houches School lecture notes, has 3 main parts. The first part describes how to construct a Hamiltonian for a general circuit, which can include dissipative elements. The second part describes the quantization of the circuit, with an emphasis on the quantum treatment of dissipation. The final part focuses on the Josephson non-linear element and the main linear building blocks from which superconducting circuits are assembled. It also includes a brief review of the main types of superconducting artificial atoms, elementary multi-level quantum systems made from basic circuit elements.}


\end{titlingpage}

\tableofcontents

\newpage
\section{What are quantum electromagnetic circuits?}

\subsection{Macroscopic quantum mechanics}

One usually associates quantum mechanics with microscopic particles such as
electrons, atoms or photons and classical mechanics with macroscopic objects
such as billiard balls, solar systems and ocean waves. In recent years
however, the notion has emerged that some systems, now referred to as
mesoscopic systems, have a status intermediate between microscopic quantum
particles and macroscopic classical objects~\cite{Caldeira1983,Leggett1987}. Like billiard
balls, they are macroscopic in the sense that they contain a large number of
atoms and are \textquotedblleft artificial\textquotedblright , i.e. they are
man-made objects designed and built according to certain specifications.
However, they also possess collective degrees of freedom, analogous to the
position of the center-of-mass of the ball, that behave
quantum-mechanically. The parameters influencing this quantum behavior are
phenomenological parameters which can be tailored by the design of the
system and not fundamental, \textquotedblleft God-given\textquotedblright\
constants like the Bohr radius or the Rydberg energy. Mesoscopic physics is
a new area of research where novel quantum phenomena that have no equivalent
in the microscopic world can be imagined and observed.

\begin{figure} [htp]
\centering
\includegraphics[angle = 0, width =7.9584cm]{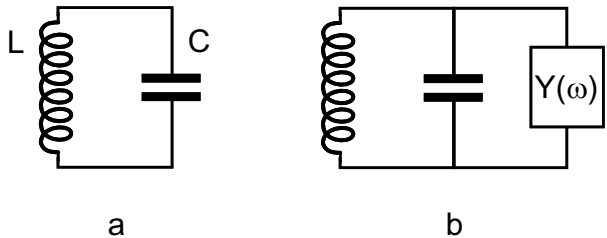}
\caption{\label{lc}  (a) Isolated ideal LC
oscillator. (b) LC oscillator connected to an electromagnetic environment
represented by an admittance $Y(\protect\omega )$ in parallel with the
circuit.}
\end{figure}

To make the discussion more concrete, let us imagine a LC oscillator circuit
(see Fig.~\ref{lc}a) fabricated with the technology of microelectronic
chips. We suppose that the oscillator is isolated from the rest of the chip
and we take internal dissipation to be vanishingly small. Typical values
that can be easily obtained for the inductance and the capacitance are $L=1~%
\mathrm{nH}$ and $C=10~\mathrm{pF}$. They lead to a resonant frequency $%
\omega _{0}/2\pi =1/2\pi \sqrt{LC}\simeq 1.6~\mathrm{GHz}$ in the microwave
range. Nevertheless, because the overall dimensions of the circuit do not
exceed a few hundred $\mathrm{\mu m}$, which is much smaller than the
wavelength corresponding to $\omega _{0}$ (around 20cm), the circuit is well
in the lumped element limit. It is described with only one collective degree
of freedom which we can take as the flux $\Phi $ in the inductor. This
variable is the convenient electrical analog of the position of the mass in
a mass-spring mechanical oscillator, the momentum of the mass corresponding
to the charge $Q$ on the capacitor. The variables $\Phi $ and $Q$ are
conjugate coordinates in the sense of Hamiltonian mechanics.

The chip on which this circuit has been patterned is enclosed in a
well-shielded copper box anchored thermally to the cold stage of a dilution
refrigerator at $T=20~\mathrm{mK}$. With these precautions, $k_{B}T\ll \hbar
\omega _{0}$, i.e. the thermal fluctuation energy is much smaller than the
energy quantum associated with the resonant frequency (this energy
corresponds to about 75mK if we express it as a temperature). But this
latter condition is not sufficient to ensure that $\Phi $ needs to be
treated as a quantum variable: the width of the energy levels must also be
smaller than their separation. This means that the quality factor of the LC
oscillator needs to satisfy $\mathcal{Q}\gg 1$, a constraint on the damping
of the oscillator.

Of course, a superconducting metal can be used for the wire of the inductor.
But we also need to make measurements on the circuit via leads which can
transfer energy in and out the oscillator. The leads and the measuring
circuit constitute the electromagnetic environment of the LC oscillator. The
strong coupling between the oscillator and its environment is the main
limiting factor for the quanticity of $\Phi $. The influence of the
environment on the oscillator can be modeled as a frequency dependent
admittance $Y\left( \omega \right) $ in parallel with the capacitance and
the inductance (see Fig. 1b). The environment shifts the oscillator
frequency by the complex quantity $\Delta +\frac{i}{2}\omega _{0}/\mathcal{Q}%
\simeq \omega _{0}\left[ \frac{i}{2}Z_{0}Y\left( \omega _{0}\right) -\frac{1%
}{8}Z_{0}^{2}Y\left( \omega _{0}\right) ^{2}-\frac{\omega _{0}}{2}%
Z_{0}^{2}Y\left( \omega _{0}\right) Y^{\prime }\left( \omega _{0}\right) %
\right] $, where $Z_{0}=\sqrt{\frac{L}{C}}$ is the impedance of the elements
of the oscillator on resonance (here we are neglecting terms of order $%
\left( Z_{0}Y\right) ^{3}$ and higher orders). In our example $Z_{0}$ has
the value 10$\Omega $. With present day technology, we can engineer a
probing circuit that would submit the oscillator to only thermal
equilibrium noise at 20mK while loading it with a typical value for $%
\left\vert Y\left( \omega \right) \right\vert ^{-1}$ in the range of 100$%
\Omega $ or above\footnote{At microwave frequencies, impedances tend to be of the order of the impedance of the vacuum $Z_{\mathrm{vac}}=\left( \mu_{0}/\epsilon _{0}\right) ^{1/2}\simeq 377~\Omega $.}. The value 100$\Omega $ corresponds to $%
\mathcal{Q}=10$. This example shows how electrical circuits, which are
intrinsically fast and flexible, constitute a class of mesoscopic quantum
systems well adapted to experimental investigations.

However, the particular LC circuit we have considered is too simple and only
displays rather trivial quantum effects. Because it belongs to the class of
harmonic oscillators, it is always in the correspondence limit. The average
value of the position or the momentum follow the classical equations of
motion. Quantum mechanics is revealed in the variation with temperature of
the variances $\left\langle \Phi ^{2}\right\rangle $ and $\left\langle
Q^{2}\right\rangle $, but these higher moments of the basic variables are
considerably much more difficult to measure than the average of these
quantities. Remember that we are dealing here with a
system possessing a single degree of freedom, instead of a thermodynamic system.

\begin{figure} [htp]
\centering
\includegraphics[angle = 0, width =7.5366cm]{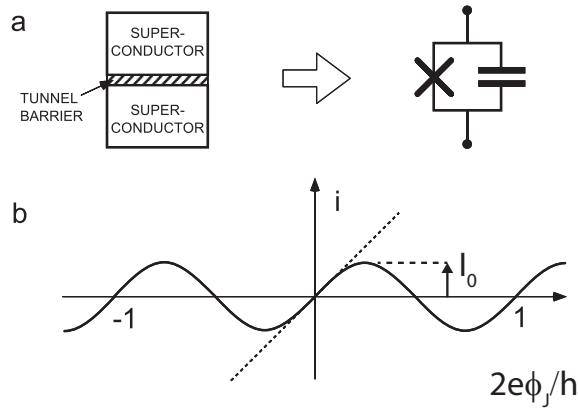}
\caption{\label{jj-basics}  (a) A Josephson tunnel junction
can be modeled as a Josephson tunnel element (cross) in parallel with a
capacitor. (b) Current-flux relation of the Josephson element. The dashed
line is the current-flux relation of a linear inductance whose value is
equal to the effective inductance of the junction. The solid line is the
relationship between the current traversing the Josephson element and the
generalized flux across it (see text).}
\end{figure}

Non-trivial and directly observable macroscopic quantum effects appear in
circuits which contain at least one non-linear component. At the time of
this writing, the Josephson tunnel junction is the best
electrical component that is sufficiently both non-linear and
non-dissipative at temperatures required for the observation of macroscopic quantum
effects~\footnote{Recent advances in the field of superconducting nanowires, nanomechanical oscillators and atomic point contacts may bring alternative elements.}. The Josephson tunnel junction consists of a
sandwich of two superconducting electrodes separated by a 1nm-thin oxide
layer (see Fig.~\ref{jj-basics}a). It is modeled electrically as a pure
superconducting tunnel element (also called Josephson element), which can be
thought of as a non-linear inductor (Fig.~\ref{jj-basics}b), in parallel
with a capacitance. The latter corresponds to the parallel plate capacitor
formed by the two superconductors. The Josephson element is traditionally
represented by a cross in circuit diagrams. The origin of the non-linearity
of the Josephson element is very fundamental: as we will see, it is
associated with the discreteness of charge that tunnels across the thin
insulating barrier.

At a temperature of a few tens of mK, all the electrons in the
superconducting electrodes on each side of the junction are condensed into
Cooper pairs. All internal degrees of freedom in the electrodes are thus
frozen and the junction is characterized only by two \textit{a priori }
independent collective degrees of freedom: the charge $Q\left( t\right) $ on
the capacitance and the number $N\left( t\right) $ of Cooper pairs having
tunneled across the Josephson element. The charge $Q_J\left( t\right)
=-2eN\left( t\right) $ having flown through the Josephson element up to a
time $t$ need not be equal to $Q\left( t\right) $ if the junction is
connected to an electrical circuit. Note that while $Q$ is a continuous
variable corresponding to a bodily displacement of the electron fluid in the
electrodes with respect to the ion lattice, $N$ is an integer variable. The
Josephson element can also be characterized by a generalized flux $\phi _J$, a
position-like variable which can be defined as the time integral of the instantaneous voltage $v_J$ across
the element.

\begin{equation}
\phi _J\left( t\right) =\int_{-\infty }^tv_J\left( t^{\prime }\right)
dt^{\prime }  \label{phiJdef}
\end{equation}

At time $t=-\infty ,$ all electromagnetic fields in the circuit are supposed
to have been zero and the voltage $v_J$ includes in particular electromotive
forces due to the appearance of magnetic field through the loops of the
circuit containing the Josephson junction. One can check that this
definition of the generalized flux agrees with the usual definition of flux
for an inductor whose leads are joined, as it then encloses a precisely
defined area through which the flux of the instantaneous magnetic field can
be evaluated.

Whereas for an inductance $L$, there is a linear relation between the current 
$i\left( t\right) $ that flows through it and the generalized flux $\phi_L
\left( t\right) $ across it

\begin{equation}
i\left( t\right) =\frac 1L\phi_L \left( t\right)  \label{constofL}
\end{equation}

the Josephson element is characterized by the following current-flux
relation:

\begin{equation}
i\left( t\right) =I_{0}\sin \left[ \frac{2e}{\hbar }\phi _{J}\left( t\right) %
\right]  \label{constofJJ}
\end{equation}%
As previously mentioned, the scale of non-linearity in this relation is set
by the superconducting flux quantum $\phi_0=\hbar /2e$ based on the Cooper pair
charge $2e$. The dimensionless combination $\varphi =2e\phi _{J}/\hbar $ is
known under the esoteric name \textquotedblleft gauge-invariant phase
difference\textquotedblright\ or simply \textquotedblleft phase
difference\textquotedblright. The presence of $\hbar $ in the argument of
the sine function in the current-flux relationship should not obscure the
fact that $\phi _{J}$ is a macroscopic collective variable involving the
electrical analog of the center of mass of all electrons in the junction.
For $|\phi _{J}|\ll \phi_0$, the tunnel element behaves as an inductance with
a value $L_{J}=\phi_0 /I_{0}$.

Josephson's unexpected discovery~\cite{Josephson1962,Josephson1969} was that the parameter $%
I_{0}$ (and correspondingly $L_{J})$ which characterizes the tunnel element
is a macroscopic parameter in the sense that it is proportional to the area
of the junction. Note it is also proportional to the transparency of the
tunnel barrier, which depends exponentially on its thickness. Typical values
for $I_{0}$ in experiments on macroscopic quantum effects are in the $\mu 
\mathrm{A}-\mathrm{nA}$ range. Correspondingly, the junction effective
inductances are in the range $\mathrm{nH}-\mathrm{\mu H,}$ while the
junction capacitances, determined by the area and thickness of the oxide
layer, are in the $\mathrm{pF}-\mathrm{fF}$ range. These orders of magnitude
make characteristic frequencies of the junction in the GHz range. There is
thus a similarity between experiments on quantum effects in Josephson
junction systems and Rydberg atom cavity QED experiments \cite{Haroche-Raimond}.
Josephson junctions play the role of Rydberg atoms while the embedding
circuit plays the role of the cavity and the preparation/detection
apparatuses. This is why in recent years the field of quantum Josephson circuits has often been nicknamed ``circuit QED" \cite{Blais2004}.

In contrast with the quantum fluctuations of the LC oscillator, which are
completely decoupled from externally imposed currents and voltages, the quantum
fluctuations of a Josephson junction (or of more complex systems involving
several junctions) manifest themselves directly in the RF response of the circuit due to the junction nonlinearity. 
This relative experimental simplicity has a
counterpart, however. Josephson junctions are so well coupled to their
electromagnetic environment that dissipation cannot always be treated as a
perturbation. In fact, dissipation combines with the nonlinearity of tunnel
elements to produce qualitatively new quantum effects which are not
encountered for example in the almost dissipation-free quantum systems
studied in atomic physics. One of the most spectacular new quantum features is the
localization of position-like degrees of freedom when dissipation exceeds a
certain threshold set by the quantum of resistance $h/(2e)^{2}\simeq 6.4~%
\mathrm{k}\Omega $~\cite{Grabert-Devoret}.

\subsection{From fields to circuits, and circuits to fields}
%
%Circuits are caricatures of fields, systems with distributed DOF. But at the same time, if a circuit has an infinite number of elements, it `s the microscopic description of a field.

\begin{figure} [htp]
\centering
\includegraphics[angle = 0, width =8.0396cm]{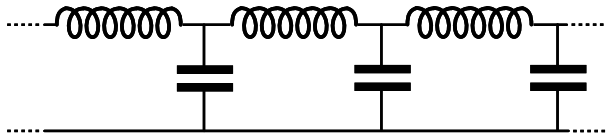}
\caption{\label{ladder-circuit}  LC ladder circuit. In the
limit of an infinite number of elements, it can model the propagation of the
TEM mode of a coaxial transmission line.}
\end{figure}

Distributed electromagnetic systems can be represented by lumped
element circuits as long as the properties of the lowest frequency modes of the system are considered.
For instance, the link between a microwave cavity and an LC oscillator is very well
discussed by Feynman~\cite{Feynman}. In this representation, inductances and capacitances can be considered as
\textquotedblleft bottles\textquotedblright\ for magnetic and electric
fields respectively. On the other hand, a circuit with an infinite number of circuit elements can be treated as a continuous electromagnetic field model. A simple example is the
infinite LC ladder with pitch $a$ (see Fig.~\ref{ladder-circuit}) which sustains
propagating modes that are equivalent, in the limit of wavelengths $\lambda \gg a$, to the TEM modes of a coaxial transmission
line. This kind of reverse correspondence is at work in the field of electromagnetic meta-materials. 
The Hamiltonian formulation is useful in the exploration of such
correspondences.

\subsection{Superconducting qubits for quantum information}

The concept of solving problems with the use of quantum algorithms, introduced in the early 1990s~\cite{NielsenChuang,Mermin}, was welcomed as a revolutionary change in the theory of computational complexity, but the feat of actually building a quantum computer was then thought to be impossible. The invention of quantum error correction (QEC)~\cite{Shor1995,Steane1996,Knill1997,GottesmanThesis} introduced hope that a quantum computer might one day be built, most likely by future generations of physicists and engineers.  However, 20 years later, we have witnessed so many advances that successful quantum computations, and other applications of quantum information processing (QIP) such as quantum simulation~\cite{Lloyd1996,Cirac2012} and long-distance quantum communication~\cite{Kimble2008}, appear reachable within our lifetime, even if many discoveries and technological innovations are still to be made.

A recent review discusses the specific physical implementation of general-purpose QIP with superconducting qubit circuits~\cite{Devoret2013}, now a major contender for the realization of a scalable quantum computer. Unlike microscopic entities  - electrons, atoms, ions, and photons - on which other qubits are based, superconducting quantum circuits are built starting from  electrical oscillators and are macroscopic systems with a large number of (usually aluminum) atoms assembled in the shape of metallic wires and plates. The operation of superconducting qubits is based on two robust phenomena: superconductivity, which is the frictionless flow of electrical fluid through the metal at low temperature (below the superconducting phase transition), and the Josephson effect, which endows the circuit with nonlinearity without introducing dissipation or dephasing. The collective motion of the electron fluid around the quantum  circuit is analogous to the position of the electron in an atom serving as qubit. The Josephson tunnel junction ensures that the circuit behaves as a true artificial atom, for which the transition from the ground state to the excited state ($\ket{g}$-$\ket{e}$) can be selectively excited and used to manipulate the qubit, unlike in the pure LC harmonic oscillator. What is remarkably rich in the implementation of a quantum processor with superconducting circuits, in addition to its realization using the techniques of integrated circuits, is the diversity of system Hamiltonians that can be designed and implemented to perform a given function. This point will be addressed in some detail in part 4 of this review. 

 \subsection{How is this article organized?} 
 
This article, which is an updated and modernized version of a previous set of Les Houches School lecture notes\cite{DevoretNotes1997}, is not intended as a comprehensive review of the now important
literature on quantum effects in tunnel junction circuits. It rather aims
at discussing some basic concepts which, in the opinion of the authors, are
important for understanding the various points of view adopted in the
specialized articles, as well as clarifying some difficult detailed points.

Thus, the references given in this review constitute an incomplete and
subjective picture of the field. They must be thought of only as entry
points in the literature. We extend our apologies to the authors of many important works which are not cited in this review.

We organized this article as follows. In the next section, we explain how
the Hamiltonian formalism, which provides a well-trodden path to go from the
classical to the quantum description of a system, can be applied to
electrical circuits. Whereas the Hamiltonian framework can be
straightforwardly applied to the LC oscillator of Fig.~\ref{lc}, it is much
less obvious to do so in complicated circuits, in particular with non-linear
elements, and we describe a systematic procedure. A thorough understanding
of the classical properties of tunnel junction circuits is needed to
clearly separate effects due to the non-linear constitutive relation of
tunnel elements (which originates from microscopic quantum effects in the
junction and can be taken as purely phenomenological) and genuine
macroscopic quantum effects originating from quantum fluctuations of
macroscopic electrical quantities. We then treat in the following section
the quantum mechanics of linear dissipative circuits. We discuss in
particular the case of the LC circuit with damping. The quantum fluctuations
of this system can be computed analytically and they provide a useful
benchmark for the quantum fluctuations and interferences in simple circuits involving
Josephson junctions which are treated in the following section. In
particular, we discuss the case of the Cooper pair box, from which many
other circuits can be derived. We finish by discussing artificial atoms involving Josephson junction arrays and more generally, the families of circuits generated by varying on one hand the ratio of Coulomb to Josephson energy, and on the other hand, the ratio of the Josephson inductance to the effective circuit inductance shunting the junction. The article ends by a short summary of previous sections and survey of the perspective of quantum circuits.

\newpage
\section{Hamiltonian description of the classical dynamics of electromagnetic circuits}
\subsection{Non-dissipative circuits}

\subsubsection{Circuit definitions}

\begin{figure} [htp]
\centering
\includegraphics[angle = 0, width =5.1489cm]{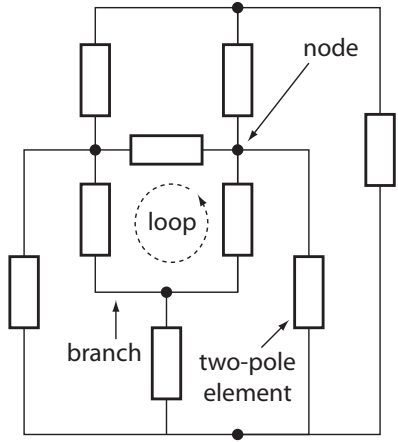}
\caption{\label{node-circuit-example}  An electrical circuit consists of two-pole elements forming the branches of the network and meeting at nodes. Loops are formed when there is more than one path between two nodes.}
\end{figure}

An electrical circuit can be formally described as a network of elements connected at nodes (See Fig.~\ref{node-circuit-example}).
With little loss of generality, we consider only two-pole elements, which are connected only to two nodes. For a more mathematically complete discussion of
networks, see Ref.~\cite{Burkard2004}. These two-pole elements
form the branches of the network. 

\subsubsection{Dynamical variables of the circuit}

\begin{figure} [htp]
\centering
\includegraphics[angle = 0, width =3.4772cm]{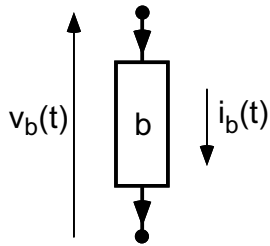}
\caption{\label{sign-convention}  Sign convention for the
voltage and current associated with an arbitrary branch $b$ of an electrical
circuit.}
\end{figure}

The element of each branch $b$ at time $t$ is characterized by two variables: the voltage $v_{b}\left( t\right)$ across the elements and the current $i_{b}\left( t\right) ]$
flowing through it (see Fig.~\ref{sign-convention}). For each branch $b$ we choose an orientation, arbitrary at this point, which will determine the sign of the current value. The voltage orientation is chosen to be opposite to that of the current for reasons that will become clear later.

The voltage and the current are defined from the underlying electromagnetic fields by

\begin{equation}
v_b(t)=\int_{\mathrm{beginning~of~}b}^{\mathrm{end~of~}b}\overrightarrow{E}(\overrightarrow{r},t)\cdot 
\overrightarrow{d\ell}  \label{vdef}
\end{equation}

\begin{equation}
i_{b}(t)=\frac{1}{\mu _{0}}\oint_{\mathrm{around~}b}\overrightarrow{B}(\overrightarrow{r},t)\cdot 
\overrightarrow{ds}  \label{idef}
\end{equation}%

In Eq.~\ref{idef} the loop integral is done along a closed curve in vacuum encircling the element.

Because we consider circuits in the lumped element approximation, these
definitions make voltages and currents independent, to a large extent, of the precise path of
integration along which fields are integrated. These paths are well outside the
wire of inductors for the line integral of electric field (so that the magnetic field is zero along the path) and well outside
the dielectric of capacitors for the loop integral of magnetic field (so the electric field is zero along the loop). Note
that these definitions are sufficiently general to include the contribution
to voltages of electro-motive forces due to time-varying magnetic fields and the
contribution to currents of displacement currents due to time-varying
electric fields. Note also that the factor $\mu _{0}$ in the definition of
current comes from our choice of working with SI units throughout this
review.

\subsubsection{Energy absorbed by an element}

The power absorbed by an element is given by the product of the voltage and current defined above (note the relevance of the sign convention here). We now introduce the total energy absorbed by an element $b$:
\begin{equation}
\mathcal{E}_b(t) = \int_{-\infty}^{t} v_b(t')i_b(t')dt'
\end{equation}

In this expression the lower bound of the integral ($t' = -\infty$) refers actually to a time sufficiently far in the past that the circuit was completely at rest (This of course assumes the circuit contains a small amount of dissipation).
An element is said to be purely  ``dispersive'' (or ``conservative'') if the energy $\mathcal{E}$ is converted into stored electric or magnetic energy.

\subsubsection{Generalized flux and charge associated with an element}

An Hamiltonian description of electrical circuits requires the introduction
of branch fluxes and branch charges which are defined from branch voltages
and branch currents by

\begin{eqnarray}
\Phi _b\left( t\right) &=&\int_{-\infty }^tv_b(t^{\prime })dt^{\prime }
\label{phidef} \\
Q_b\left( t\right) &=&\int_{-\infty }^ti_b(t^{\prime })dt^{\prime }
\label{qdef}
\end{eqnarray}

As stated previously, the circuit is supposed to have been at rest at time $t=-\infty $ with zero
voltages and currents. Static bias fields imposed externally on the circuit
such as magnetic fields through the inductors are supposed to have been
switched on adiabatically from $t=-\infty $ to the present.

\subsubsection{Capacitive and inductive elements}

A dispersive element for which the voltage $v(t)$ is only a function of the charge $Q(t)$ and not directly of the time $t$ or any other variables, is said to be a capacitive element. 
\begin{equation}
v(t)=f(Q(t))  \label{capaconst}
\end{equation}

Its capacitance, which is only a function of the charge, is given by:
\begin{equation}
C(Q) = \left[\frac{df}{dQ}\right] ^{-1}  
\end{equation}

A linear capacitance has $C(Q) = C$ independent of $Q$ and $v(t) = (Q(t)-Q_\mathrm{offset})/C$. One can easily compute that in this case $\mathcal{E}(t) = \frac{1}{2C}(Q(t)-Q_\mathrm{offset})^2$.

Similarly, a dispersive element for which the current $i(t)$ is only a function of the flux $\Phi(t)$ and not directly of the time $t$ or any other variables, is said to be an inductive element. 
\begin{equation}
i(t)=g(\Phi(t))  \label{selfconst}
\end{equation}

Its inductance, which is only a function of the flux, is given by:
\begin{equation}
L(\Phi) = \left[\frac{dg}{d\Phi}\right]^{-1}  
\end{equation}

A linear inductance has $L(\Phi) = L$ independent of $\Phi$ and $i(t) = (\Phi(t)-\Phi_\mathrm{offset})/L$. One can easily compute that in this case $\mathcal{E}(t) = \frac{1}{2L}(\Phi(t)-\Phi_\mathrm{offset})^2$.

As we have seen with Eq.~\ref{constofJJ}, a Josephson tunnel junction possess a non-linear inductive
element for which $g$ is a sine function: $i(t)=I_{0}\sin \left( 2e(\Phi(t)-\Phi_\mathrm{offset})
/\hbar \right) $.

In summary,  the energies of our three basic elements are given by the following table:

\begin{center}
\begin{tabular}{|c|c|}
\cline{1-2}
Element & Energy \\ 
\hhline{|==|}
&\\[-1em]
linear capacitance $C$ & $\frac{1}{2C}(Q-Q_\mathrm{offset})^2$ \\ 
&\\[-1em]
\cline{1-2}
&\\[-1em]
linear inductance $L$ & $\frac{1}{2L}(\Phi-\Phi_\mathrm{offset})^2$ \\ 
&\\[-1em]
\cline{1-2}
&\\[-1em]
Josephson element $L_J$ & $\frac{\phi^2_0}{L_J}\left[ 1-\cos \left(
(\Phi-\Phi_\mathrm{offset})/\phi_0 \right) \right] $\\
\multicolumn{1}{c}{}&\multicolumn{1}{c}{}\\[-1em]
\cline{1-2}
\end{tabular}
\end{center}

Let us stress that despite the presence of $\hbar $ and $e$ in the
expression of the energy of the Josephson element (through $L_J$ and $\phi_0$), it is at this stage a
purely classical entity, which, from the point of view of collective
variables like current and voltages, is on the same footing as a common
inductance obtained by winding a piece of macroscopic wire. Universal
quantum constants enter here only because the non-linear behavior of this
element originates from the microscopic phenomenon of discrete electron
tunneling events between the electrodes.
Equations like Eqs.~\ref{capaconst} and~\ref{selfconst} are called the constitutive equations of the element.

\begin{figure} [htp]
\centering
\includegraphics[angle = 0, width =5.1489cm]{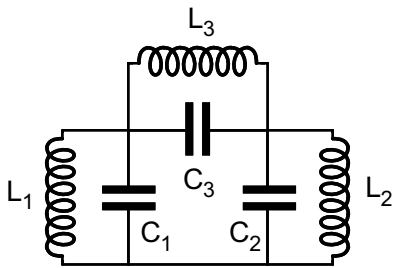}
\caption{\label{generic-circuit}  Example of non-dissipative
circuit whose branches consist of linear inductances and capacitances. The
nature and number of degrees of freedom of the circuit would not change if
the linear elements were replaced by non-linear ones.}
\end{figure}

A linear dispersive circuit consists only of linear capacitances and inductances, for example see Fig.~\ref{generic-circuit}.

\subsubsection{Finding the degrees of freedom of an arbitrary conservative circuit}

We suppose the circuit is sufficiently near rest that the constitutive equations can be linearized, i.e. the energy can be 
expanded as a quadratic term plus higher order corrections. The problem is now reduced to finding the degrees of freedom of the linear dispersive circuit corresponding to the quadratic term.

There are less degrees of freedom than there are branches in the circuits, since in addition to the constitutive relations, one has to take in to account Kirchhoff's laws:
\begin{eqnarray}
\sum_{\mathrm{all~}b~\mathrm{around~}l}\Phi _{b} &=&\widetilde{\Phi }_{l}
\label{Kirphi} \\
\sum_{\substack{ \mathrm{all~}b~\mathrm{arriving}  \\ \mathrm{~at~}n}}Q_{b}
&=&\widetilde{Q}_{n}  \label{Kirq}
\end{eqnarray}%

One therefore has to eliminate superfluous variables. There exist two standard methods in circuit theory to achieve this goal: the method of nodes and the method of loops. Here we develop only the method of nodes which solves most practical problems. The two methods are dual to each other and the lessons learned in studying one of the them are easily transposed to the other.

Before examining the details of the method of nodes, one should first mentally divide the circuit into its capacitive sub-network and inductive sub-network. In the method of nodes we turn our attention away from the loops, which pose no problems, to face what happens at a node. \textit{Active} nodes are defined as nodes in which inductances and capacitances meet. \textit{Passive} nodes are where only capacitances or only inductances converge.

\subsubsection{Method of nodes}

In the method of nodes, we exploit the specificity of the capacitive sub-network to contain only linear elements. This is a reasonable assumption for the circuits we will be discussing. This assumption allows us to express the energy of a capacitance in terms of voltage, i.e. the derivative of flux. Inverting the constitutive relation given in Eq.~\ref{capaconst}, we can write the energy of a capacitive branch as $\mathcal{E} = \frac{C}{2}\dot\phi^2$. Thus in our treatment, we have broken the symmetry between charge and flux, and flux will play the role of ``position''. With this choice inductive energy will be potential energy and capacitive energy will be kinetic energy.

We now proceed by explaining the technical details of the method of nodes. One first makes sure that at every node to which an inductance is connected, a capacitance is also connected. This does not need to be an artificial introduction, it corresponds to the always present parasitic capacitance of inductances. There are thus no passive nodes in the sub-network of inductances. On the other hand, it does not matter if this sub-network is not simply connected.
In contrast, for the capacitive sub-network, we have to make sure it is simply connected. It can, however, have passive nodes.
Thus, along with the symmetry between charges and fluxes, the symmetry between capacitances and inductances is broken in the method of nodes.
We have thus ensured that every node is connected to any other node by a path involving only capacitances.

Listing all the nodes, the active nodes will be nodes $1$ to $N$, while nodes $N+1$ to $P$ will be passive nodes of the capacitance sub-network. We first setup the $P\times P$ inverse inductance matrix $[L^{-1}]_{jk}$ whose non-diagonal matrix elements are $-1/L_{jk}$ where $L_{jk}$ is the value of the inductance connecting nodes $j$ and $k$. Of course if there is no inductance between the nodes, which is true in particular for all passive nodes, the corresponding matrix element will be zero. The diagonal matrix elements will be the opposite of the sum of values in the corresponding row or column.
We also introduce the $P\times P$ capacitance matrix $[C]_{rs}$ whose non-diagonal matrix elements are $-C_{rs}$ where $C_{rs}$ is the capacitance connecting nodes $r$ and $s$. The diagonal elements of the capacitance matrix are built similarly to that of the inductance matrix, by taking the opposite of the sum of values in the corresponding row or column.

\begin{figure} [htp]
\centering
\includegraphics[angle = 0, width =3.4772cm]{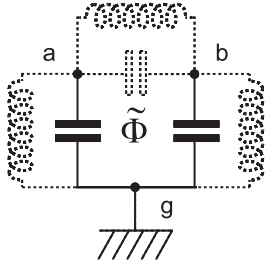}
\caption{\label{spanning-tree}  Example of spanning tree for the
circuit of Fig. \protect\ref{generic-circuit}. The ground is indicated by a
rake-like symbol. Closure branches are in dashed line. The constant $%
\widetilde{\Phi }$ is the magnetic flux through the loop formed by the three
inductors.}
\end{figure}

What is the meaning of these matrices?  Let us introduce the spanning tree of the capacitance sub-network. It consists of the choice of a particular active node called ``ground'' and the set of branches that connect the ground through capacitances to every other node, both active and passive, without forming any loops. There is thus only one path between the ground and every other node. For an example, see Fig.~\ref{spanning-tree}. This spanning tree allows us to assign a flux to each node by algebraically summing all the fluxes of the branches in the path between the ground and the node. We can now define the node flux column vector $\overrightarrow{\phi}$ which has $P-1$ components. 
The choice of ground node and spanning tree is analogous to the choice of a particular gauge in electromagnetic field theory and to the choice of a system of position coordinates in classical mechanics.

These node flux are related to the branch fluxes by the relation:
\begin{eqnarray}
\Phi _{b\in \mathsf{T}} &=&\phi _{n}-\phi _{n^{\prime }}  \label{b-ntree} \\
\Phi _{b\in \mathsf{\bar{T}}} &=&\phi _{n}-\phi _{n^{\prime }}+\widetilde{\Phi}%
_b  \label{b-nclos}
\end{eqnarray}%
where $T$ is the set of spanning tree branches and $\bar{T}$ is the complement of this set. The symbols $n$ and $n^\prime$ denote the nodes connected by the branch. The flux offset $\widetilde{\Phi}_b$
corresponds to the static flux that may be enclosed by a loop containing the branch.

We have defined the energy for each branch, and now we can express it using node fluxes. This leads to the equivalent of the potential energy of the circuit:
\begin{equation}
\mathcal{E}_\mathrm{pot} = \frac{1}{2}\overrightarrow{\phi}^t [\bm{L^{-1}}] \overrightarrow{\phi} + \sum_b \frac{1}{L_b} (\phi _{n}-\phi _{n^{\prime }}) \widetilde{\Phi}_b
\end{equation}
where the matrix $[\bm{L^{-1}}]$ differs from $[L^{-1}]$ in that the row and column corresponding to the ground node have been eliminated.
The second term sums over all the inductive branches of the circuit where $n$ and $n^\prime$ are the nodes connected by branch $b$. If there is no inductor the term will be zero. The offset fluxes represent a strain of the inductance sub-network that is crucial to a wide range of phenomena involving Josephson junctions~\cite{Likharev,VanDuzer1999,BaronePaterno}. We will describe some of these effects later in the review. 

The equivalent to the kinetic energy of the circuit is given by:
\begin{equation}
\mathcal{E}_\mathrm{kin} = \frac{1}{2}\overrightarrow{\dot\phi}^t [\bm{C}] \overrightarrow{\dot\phi} 
\end{equation}

where the matrix $[\bm{C}]$ differs from $[C]$ in that the row and column corresponding to the ground node have been eliminated.
Note that there are no offsets in the kinetic energy term due to our choice of the tree to pass only through capacitances. The charge offset will appear later when we define the conjugate charges of node fluxes.

\subsubsection{Setting up the Lagrangian}

We can now obtain the Lagrangian by subtracting the potential energy from the kinetic energy $\mathcal{L} = \mathcal{E}_\mathrm{kin}-\mathcal{E}_\mathrm{pot}$. 
For the circuit of Fig.~\ref{generic-circuit} and the choice of spanning tree of Fig.~\ref{spanning-tree}, one obtains the Lagrangian

\begin{eqnarray}
\mathcal{L}\left( \phi _a,\dot{\phi}_a,\phi _b,\dot{\phi}_b\right) &=&\frac{%
C_1\dot{\phi}_a^2}2+\frac{C_2\dot{\phi}_b^2}2+\frac{C_3\left( \dot{\phi}_a-%
\dot{\phi}_b\right) ^2}2  \notag \\
&&-\left[ \frac{\phi _a^2}{2L_1}+\frac{\phi _b^2}{2L_2}+\frac{\left( \phi
_a-\phi _b+\widetilde{\Phi }\right) ^2}{2L_3}\right]  \label{lagrangdef}
\end{eqnarray}
where the degrees of freedom $\phi _{a}$ and $\phi _{b}$ are the fluxes of
the nodes $a$ and $b$. One can check that by applying Lagrange's equations

\begin{equation*}
\frac{d}{dt}\frac{\partial \mathcal{L}}{\partial \dot{\phi}_{n}}-\frac{%
\partial \mathcal{L}}{\partial \phi _{n}}=0
\end{equation*}
one recovers the correct equations of motion of the circuit. Our approach for the
construction of the Lagrangian of a circuit generalizes the pioneering work of Yurke
and Denker~\cite{Yurke1984}.

\subsubsection{Conjugate variable pairs}

From the Lagrangian, we can now define the momenta conjugate to the node
fluxes, using the usual relation

\begin{equation}
q_n=\frac{\partial \mathcal{L}}{\partial \dot{\phi}_n}  \label{conjdef}
\end{equation}

It is important to note that, according to Lagrange's equations, $\dot{q}_n = 0$ if $n$ is a passive node, since by our definition $\frac{\partial \mathcal{L}}{\partial {\phi}_n} = 0$ for all passive nodes in the capacitive sub-network. Thus, the circuit only has at most $N-1$ true degrees of freedom, corresponding to all the active nodes except the ground node.
These new variables $q_{n}$, which we call node charges, correspond to the algebraic sum of the charges on the capacitances connected to node $n$. In
the loop variable representation, the conjugate momentum of the loop charge is the sum of the fluxes in the inductors of the loop. 

Note that Eq.~\ref{conjdef} can be written in vector form as $\overrightarrow{q} = [\bm{C}] \overrightarrow{\dot\phi}$. It is possible to invert the capacitance matrix and thus express $\overrightarrow{\dot\phi}$ as a function of $\overrightarrow{q}$. 

We can now find the normal modes of the circuit given by the eigenvectors of the matrix product $[\Omega^2]=[\bm{C^{-1}}][\bm{L^{-1}}]$ associated with non-zero eigenvalues. The non-zero eigenvalues correspond to the normal mode frequencies of the circuit squared. There are thus at most $N-1$ normal mode, but symmetries in the circuit can reduce this number. We define the number of normal modes as $M$. This number is equivalent to the number of independent equations generated by the Euler-Lagrange equations.

\subsubsection{Finding the Hamiltonian of a circuit}

The Hamiltonian can now be expressed as the sum of the kinetic energy, which is to be expressed in terms of the $q_n$ variable, and the potential energy expressed, as before, in terms of $\phi_n$:
\begin{equation}
\mathcal{H} = \frac{1}{2}\overrightarrow{q}^t [\bm{C^{-1}}] \overrightarrow{q} + \mathcal{E}_\mathrm{pot}
\end{equation}
where the independent variables $q_n$ correspond to degrees of freedom, while the others correspond to offset charges. The potential energy $\mathcal{E}_p$ is in general a non-linear function of the vector $\overrightarrow{\phi}$.

Taking again the example of the circuit of Fig.~\ref{generic-circuit}, we
can apply this procedure and obtain the following Hamiltonian

\begin{eqnarray}
\mathcal{H}\left( \phi _{a},q_{a},\phi _{b},q_{b}\right) &=&\frac{1}{%
C_{1}C_{2}+C_{1}C_{3}+C_{2}C_{3}}\left[ \frac{\left( C_{2}+C_{3}\right)
q_{a}^{2}}{2}\right.  \notag \\
&&\left. +\frac{\left( C_{1}+C_{3}\right) q_{b}^{2}}{2}+C_{3}q_{a}q_{b}%
\right]  \notag \\
&&+\left[ \frac{\phi _{a}^{2}}{2L_{1}}+\frac{\phi _{b}^{2}}{2L_{2}}+\frac{%
\left( \phi _{a}-\phi _{b}+\widetilde{\Phi }\right) ^{2}}{2L_{3}}\right]
\label{Ham}
\end{eqnarray}

The first term in $\mathcal{H}$ is the electrostatic energy of the circuit
expressed as a function of the node charges, while the second term is the
magnetic energy expressed as a function of node fluxes. This structure is a
general characteristic of the Hamiltonian of a circuit in the node variable
representation and does not depend on whether the elements are linear or
not. The Hamiltonian formulation shows clearly the role of $\widetilde{\Phi }
$ as an offset term in the magnetic energy. In the case of a linear
inductor, the effect of this term is simply to induce an offset DC current.
However, in the case of non-linear inductors like Josephson junctions, this
term changes the dynamics of the circuit.

One can easily verify that Hamilton's equations

\begin{eqnarray}
\dot{\phi}_n &=&\frac{\partial \mathcal{H}}{\partial q_n}  \label{HEM1} \\
\dot{q}_n &=&-\frac{\partial \mathcal{H}}{\partial \phi _n}  \label{HEM2}
\end{eqnarray}

are equivalent to the equations of motion.

It is important to note that although the Hamiltonian of the circuit always
gives its total energy, its functional form depends on the particular choice
of spanning tree, even when the choice of a representation in terms of node
variables or loop variables has been made.

However, the Poisson bracket~\cite{Goldstein1980} of the flux and charge of a
branch is independent of the choice of the spanning tree and obeys:

\begin{equation}
\left\{ \Phi _b,Q_b\right\} =\sum_n\frac{\partial \Phi _b}{\partial \phi _n}%
\frac{\partial Q_b}{\partial q_n}-\frac{\partial Q_b}{\partial \phi _n}\frac{%
\partial \Phi _b}{\partial q_n}=\pm1  \label{P.B.}
\end{equation}
where the value is $+1$ for a capacitance and $-1$ for an inductance. This important remark is far-reaching in the quantum case.

\subsubsection{Mechanical analog of a circuit, does it always exist?}

In the node variable representation, the node fluxes play the role of
position coordinates and the node charges the role of momentum coordinates.
The capacitive energy plays the role of the kinetic energy and the
inductive energy plays the role of the potential energy. However, the form of
the Hamiltonian of Eq.~\ref{Ham} with capacitive cross-terms shows that
the particular circuit of Fig.~\ref{generic-circuit} has no simple
mechanical analog. In the cases where the capacitances are only connected
between the active nodes and ground, they can be interpreted as the masses
of the active nodes and a direct mechanical analog can be found for the
circuit. The inductances then correspond to elastic coupling interactions
between the masses associated with the nodes. 

\subsubsection{Generalization to non-linear circuits}

It is remarkable that the formalism given above can be kept essentially intact when one goes to inductive elements with a polynomial expansion in branch fluxes. Special care must be taken, though, with the Josephson energy that is periodic in generalized flux. Also, when we deal with non-linear circuits, it is important that the capacitive sub-network remains linear. In the case of a linear inductive sub-network and a non-linear capacitive sub-network, we could resort to the method of loops. The case in which both inductive and capacitive sub-network are maximally non-linear is a subject of ongoing research.   

\subsection{Circuits with linear dissipative elements}

\subsubsection{The Caldeira-Leggett model}

We would like now to treat circuits with linear dissipative elements like
resistors. It would seem that the Hamiltonian formalism is powerless to
treat a dissipative system, whose behavior is irreversible, since Hamilton's
equations of motion, Eqs.~\ref{HEM1} and~\ref{HEM2} are invariant upon time
reversal. However, this reversibility problem can be solved by extending the formalism. This extension has in fact been made recurrently
throughout the history of theoretical physics. We will present here a
particular clear and useful version known as the Caldeira-Leggett model~\cite{Caldeira1983}
 which applies to systems with linear dissipation.

\begin{figure} [htp]
\centering
\includegraphics[angle = 0, width =3.1989in]{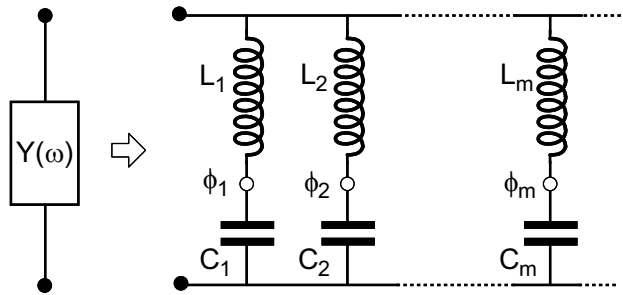}
\caption{\label{Cal-Leg}  Caldeira-Leggett model of an
admittance $Y(\protect\omega )$: the corresponding element can be
represented as an infinite number of elementary series LC circuits in
parallel. The distribution of values for the inductances and capacitances is
determined by the functional form of $Y(\protect\omega )$.}
\end{figure}

The essence of the Caldeira-Leggett model, in the context of
electrical circuits, is to replace a linear dissipative two-pole characterized by a
frequency dependent admittance $Y\left( \omega \right) $ by an infinite set
of series LC oscillators all wired in parallel (see Fig.~\ref{Cal-Leg}). The
internal degrees of freedom of the admittance can be thought of as the
fluxes of the intermediate nodes of the LC oscillators (open dots in Fig. %
\ref{Cal-Leg}). It is the passage from a finite number of degrees of freedom
to an infinite one that reconciles the irreversible behavior on physical
time scales and the formal reversibility of Hamilton's equations of motion.

\begin{figure} [htp]

\includegraphics[angle = 0, width = 6.1cm]{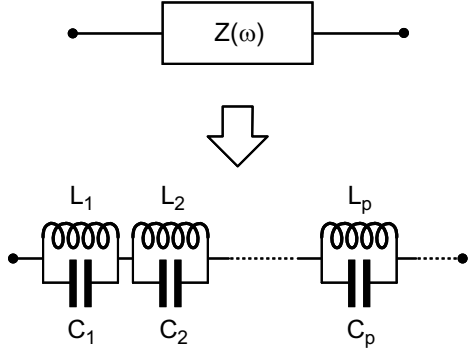}
\centering
\caption{\label{Cal-Leg-Z}  Caldeira-Leggett representation
of an impedance $Z\left( \protect\omega \right) $: the corresponding element
can be represented as an infinite set of parallel LC circuits all in series. }
\end{figure}

The reversibility problem appears when one notices that for every oscillator 
$m$ in the series, the admittance given by the usual combinatorial rules of
circuit theory

\begin{equation}
Y_m(\omega )=\left[ jL_m\omega +\frac 1{jC_m\omega }\right] ^{-1}  \label{Ym}
\end{equation}
is purely imaginary while the infinite series corresponding to $Y(\omega )$
has both a real and imaginary part (we use here the symbol

\begin{equation}
j=-\sqrt{-1}=-i  \label{EE_QM}
\end{equation}

of electrical engineers but with an opposite value to ensure later compatibility
with the sign convention of quantum mechanics concerning Fourier
transforms). This manifestation of the reversibility problem disappears by
extending the notion of admittance function to complex frequencies.

Let us recall that $Y\left( \omega \right) $ is defined from the
relationship between the voltage across a linear element and the current
flowing across it

\begin{equation}
i(t)=\int_{-\infty }^{+\infty }dt^{\prime }\ \widetilde{Y}\left( t^{\prime
}\right) v(t-t^{\prime })  \label{ZtildDef}
\end{equation}

\begin{equation}
Y\left( \omega \right) =\int_{-\infty }^{+\infty }dt\ \widetilde{Y}\left(
t\right) \exp \left( i\omega t\right)  \label{FT}
\end{equation}

We can define an extension of $Y\left( \omega \right) $ by the relation

\begin{equation}
Y\left[ \omega +i\eta \right] =\int_{-\infty }^{+\infty }dt\ \widetilde{Y}%
\left( t\right) \exp \left[ i\left( \omega +i\eta \right) t\right]
\label{fFT}
\end{equation}
(there is no problem at $t\rightarrow -\infty $ since $\widetilde{Y}\left(
t\right) $ is a causal function).

All information on the shape of $\widetilde{Y}\left( t\right) $ after $t\sim
\eta ^{-1}$ is erased in $Y\left[ \omega +i\eta \right] $. Let us now define
the generalized admittance function by

\begin{equation}
Y\left[ \omega \right] =\underset{ 
\begin{array}{c}
\eta \rightarrow 0 \\ 
\eta >0%
\end{array}
}{\lim }Y\left[ \omega +i\eta \right]  \label{ZasGreen}
\end{equation}

We find that the generalized admittance of the m-th LC circuit in the
Caldeira-Leggett model is given by

\begin{eqnarray}
Y_m\left[ \omega \right] &=&y_m\left\{ \frac \pi 2\omega _m\left[ \delta
(\omega -\omega _m)+\delta (\omega +\omega _m)\right] \right.  \notag
 \\
&&\left. +\frac i2\left[ \mathrm{p.p.}\left( \frac{\omega _m}{\omega -\omega
_m}\right) +\mathrm{p.p.}\left( \frac{\omega _m}{\omega +\omega _m}\right) %
\right] \right\}  \label{YmGen}
\end{eqnarray}
where $\omega _{m}=1/\sqrt{L_{m}C_{m}}$ and $y_{m}=\sqrt{C_{m}/L_{m}}$ are
the resonant frequency and impedance of the m-th oscillator. It has both a
real and a imaginary part. The idea of Caldeira and Leggett thus consists in
replacing the smooth $\func{Re}\left[ Y\left( \omega \right) \right] $
function by an infinitely dense comb of $\delta $ functions. Mathematically
this corresponds to the following relations between $Y\left( \omega \right) $
and the series of oscillators with finite frequency:

\begin{eqnarray}
\omega _{m\neq 0} &=&m\Delta \omega  \label{OscDescr1} \\
y_{m\neq 0} &=& \frac{2\Delta \omega}{\pi \omega_m}\func{Re}\left[ Y\left( m\Delta \omega \right) \right]\label{OscDescr2} \\
C_{m\neq 0} &=&\frac{y_{m}}{\omega _{m}}  = \frac{2\Delta \omega}{\pi \omega^2_m}\func{Re}\left[ Y\left( m\Delta \omega \right) \right]\label{OscDescr3} \\
L_{m\neq 0} &=&\frac{1}{y_{m}\omega _{m}}  =\frac{\pi}{2 \Delta \omega \func{Re}\left[ Y\left( m\Delta \omega \right) \right]}\label{OscDescr6}
\end{eqnarray}

Note that if the admittance $Y\left(\omega\right)$ corresponds to a pure conductance, all the $L_{m\neq 0}$ elements have the same value.
In order to properly treat the response of the admittance at zero frequency,
we have to introduce a 0-th element consisting only of an inductance $L_{0}$%
, with the conjugate capacitance being reduced to a short circuit ($%
C_{m=0}\rightarrow \infty $).

\begin{equation}
L_{0}=\frac{1}{\underset{\omega \rightarrow 0}{\lim }j\omega Y\left( \omega
\right) }  \label{OscDescL}
\end{equation}

Knowing the infinite set of elements, the full admittance function can be
expressed as

\begin{equation}
Y\left[ \omega +i\eta \right] =\frac{i}{L_{0}\left( \omega +i\eta \right) }+%
\underset{\Delta \omega \rightarrow 0}{\lim }\sum_{m=1}^{\infty }\left[
jL_{m}\left( \omega +i\eta \right) +\frac{1}{jC_{m}\left( \omega +i\eta
\right) }\right] ^{-1}\quad ;\quad \eta >0  \label{OscDescr}
\end{equation}

It is important to note that the Caldeira-Leggett model does not constitute
a representation of the internal workings of a dissipative element. It
should be used only to calculate the influence that such an element will
have in the dynamics of the collective variables of the circuit. We
calculate this influence by adding to the Hamiltonian of the rest of the
circuit the Hamiltonian $\mathcal{H}_{Y}$ of the admittance

\begin{equation}
\mathcal{H}_Y=\sum_m\left[ \frac{q_m^2}{2C_m}+\frac{\left( \phi_m-\phi
\right) ^2}{2L_m}\right]  \label{Hamy}
\end{equation}

This Hamiltonian has been written in the node representation where the
ground has been chosen on one terminal of the admittance. The node flux $%
\phi $ corresponds to the other terminal of the admittance while the node
fluxes $\phi _{m}$ correspond to the intermediate nodes of the LC
oscillators. The charge $q_{m}$ on the capacitances $C_{m}$ are the momenta
conjugate to $\phi _{m}$. It is useful to note that the coupling between the admittance and the circuit it is connected to is of the gauge form: the coupling term in implicitly contained in the displacement of $\phi_m$ by the main flux $\phi$.

\paragraph{Voltage and current sources}

Constant sources of voltages and current can also be treated by the Hamiltonian
formalism. A voltage source $V$ can be represented as a divergingly large
capacitor $C_{S}$ in which is stored initially a large charge $Q_{S}$ such
that $Q_{S}/C_{S}=V$ in the limit $C_{S}\rightarrow \infty $. Likewise, a
current source $I$ can be seen represented by a divergingly large inductor $%
L_{S}$ in which is stored initially a large flux $\Phi _{S}$ such that $\Phi
_{S}/L_{S}=I$ in the limit $L_{S}\rightarrow \infty $. 
Alternating voltage and current sources can, in the same manner, be treated using pre-excited LC oscillators.

\subsubsection{Fluctuation-dissipation theorem}

The value of the Caldeira-Leggett model becomes apparent when we use it to
derive the fluctuation-dissipation theorem. Suppose that the admittance $%
Y\left( \omega \right) $, which we suppose in thermal equilibrium at
temperature $T$, is short-circuited. In that case the variable $\phi $ in
the Hamiltonian (\ref{Hamy}) is identically zero and all the oscillators
become independent. The current $i\left( t\right) $ through the short is
zero on average but will fluctuate. We can easily calculate the spectral
density of these fluctuations by setting to $\frac{1}{2}k_{B}T$ the value of
each energy term in the Hamiltonian (\ref{Hamy}). For each oscillator $m$ we
can obtain the correlation function of the charge on the capacitance $C_{m}$

\begin{equation}
\left\langle q_m(t)q_m(0)\right\rangle =C_mk_BT\cos \left( \omega _mt\right)
\label{LCVCorr}
\end{equation}

The correlation function of the current through the $mth$ oscillator is
therefore

\begin{equation}
\left\langle i_m(t)i_m(0)\right\rangle =-\frac{d^2}{dt^2}\left\langle
q_m(t)q_m(0)\right\rangle =y_m\omega _mk_BT\cos \left( \omega _mt\right)
\label{LCicorr}
\end{equation}

Using the relation in Eq.~\ref{YmGen} we can rewrite this relation as

\begin{equation}
\left\langle i_m(t)i_m(0)\right\rangle =\frac{k_BT}\pi \int d\omega \func{Re}%
\left( Y_m\left[ \omega \right] \right) \exp \left( -i\omega t\right)
\label{LCPhiCorr_Z}
\end{equation}

Since all the oscillators are independent, we can add their correlation
functions to obtain the correlation of the current through the short

\begin{equation}
\left\langle i(t)i(0)\right\rangle =\sum_{m}\left\langle
i_{m}(t)i_{m}(0)\right\rangle
\end{equation}
and thus
\begin{equation}
\left\langle i(t)i(0)\right\rangle =\frac{k_{B}T}{\pi }\int d\omega \func{Re}%
\left( Y\left[ \omega \right] \right) \exp \left( -i\omega t\right)
\end{equation}

We finally obtain the spectral density of current fluctuations in
equilibrium defined by

\begin{equation}
S_I\left( \omega \right) =\int d\omega \left\langle i(t)i(0)\right\rangle
\exp \left( i\omega t\right)  \label{SVDef}
\end{equation}
in terms of the impedance function (Nyquist theorem)

\begin{equation}
S_I=2k_BT\func{Re}\left( Y\left[ \omega \right] \right)  \label{Nyquist}
\end{equation}

The spectral density of thermal equilibrium voltage fluctuations across a
linear dissipative element can be obtained as a function of its impedance $%
Z(\omega )=\left[ Y\left( \omega \right) \right] ^{-1}$ in a similar manner.
Using the Caldeira-Leggett representation of an impedance (see Fig. \ref%
{Cal-Leg-Z})

\begin{equation}
S_V=2k_BT\func{Re}\left( Z\left[ \omega \right] \right)  \label{Johnson}
\end{equation}

We will see in the next section how the quantum treatment of dissipation
modifies the results in Eqs.~\ref{Nyquist} and~\ref{Johnson}.

\newpage
\section{Hamiltonian Description of the Quantum Dynamics of Electromagnetic Circuits}

\subsection{Non-dissipative quantum circuits}

\subsubsection{From variables to operators}

The passage from the classical to the quantum description of electrical
circuit is straightforward in the framework of the Hamiltonian description
developed in the preceding section. The classical variables are replaced by
corresponding operators and the Hamiltonian function is replaced by a
function of operators:

\begin{eqnarray}
\phi &\rightarrow &\widehat{\phi }  \nonumber \\
q &\rightarrow &\widehat{q}  \nonumber \\
\mathcal{H} &\rightarrow &\widehat{\mathcal{H}}  \label{class-quant}
\end{eqnarray}

The state of the circuit is likewise represented by the density operator, which lives in the Hilbert space dual to that of the Hamiltonian.

\subsubsection{Commutators of charge and flux}

The operators corresponding to the position coordinates, here node fluxes, all commute. 
However, pairs of operators corresponding to conjugate variables do not commute. In the node variable
framework, the commutator of the node fluxes and their conjugate node charges is:
\begin{equation}
\left[ \widehat{\phi }_n,\widehat{q}_n\right] =i\hbar  \label{comm}
\end{equation}

This relation stems from the quantization of the electromagnetic field and corresponds to the fundamental commutator for conjugate variables. Of course, Eq.~\ref{comm} is valid only if the electric state of node $n$ is a true degree of freedom of the circuit, meaning that neither $\phi_n$, $q_n$ or their derivatives are constants of motion.
More generally, as shown by Dirac~\cite{Dirac}, the value of a
classical Poisson bracket imposes the value of the corresponding commutator

\begin{equation}
\left\{ A,B\right\} \rightarrow \frac 1{i\hbar }\left[ \widehat{A},\widehat{B%
}\right]  \label{clPQ2qPB}
\end{equation}

It follows from Eq.~\ref{P.B.} that the flux and the charge of a branch have the commutator

\begin{equation}
\left[ \widehat{\Phi }_b,\widehat{Q}_b\right] =\pm i\hbar  \label{commbranch}
\end{equation}
where the sign depends on the branch being capacitive or inductive.

Note, however, that in general these branch operators are not conjugate operators in the Hamiltonian. This stresses the importance of finding the correct degrees of freedom of the circuit, which can then be quantized.

\subsubsection{Useful relations}

Usual relations of quantum mechanics can be adapted to electrical systems.
For an arbitrary operator $\widehat{A}$ we have:

\begin{eqnarray}
\frac{\partial \widehat{A}/\partial \widehat{\phi }_n}{\left[ \widehat{A},%
\widehat{q}_n\right] } &=&\frac 1{i\hbar }  \label{der-comphi} \\
\frac{\partial \widehat{A}/\partial \widehat{q}_n}{\left[ \widehat{A},%
\widehat{\phi }_n\right] } &=&\frac{-1}{i\hbar }  \label{der-comq} \\
\frac{\partial \widehat{A}/\partial t}{\left[ \widehat{A},\widehat{\mathcal{H%
}}\right] } &=&\frac 1{i\hbar }  \label{der-comt}
\end{eqnarray}

The sign of the right-hand side in these relations can be obtained by
matching the order of the variables on the left-hand side to the order of
variables in the columns of the following mnemonic table:

\[
\left| 
\begin{tabular}{ccc}
$\widehat{\phi }_n$ & $\longleftrightarrow $ & $t$ \\ 
$\downarrow $ &  & $\downarrow $ \\ 
$\widehat{q}_n$ & $\longleftrightarrow $ & $\widehat{\mathcal{H}}$%
\end{tabular}
\right| 
\]

The integral form of these relations will also be useful:

\begin{equation}
A\left( t\right) =\mathrm{e}^{\frac{i\widehat{\mathcal{H}}t}\hbar }A\left(
0\right) \mathrm{e}^{-\frac{i\widehat{\mathcal{H}}t}\hbar }  \label{Heisen}
\end{equation}

\begin{eqnarray}
\mathrm{e}^{\frac{i\widehat{\phi }_nq}\hbar }\widehat{q}_n\mathrm{e}^{-\frac{%
i\widehat{\phi }_nq}\hbar } &=&\widehat{q}_n-q \\
\mathrm{e}^{\frac{i\widehat{q}_n\phi }\hbar }\widehat{\phi }_n\mathrm{e}^{-%
\frac{i\widehat{q}_n\phi }\hbar } &=&\widehat{\phi }_n+\phi  \label{trans}
\end{eqnarray}

In order to simplify notations, the hats on operators will be dropped from now on. We will of course make sure that the distinction between operators and c-numbers can be made from the context.

\subsubsection{Representations of the Hamiltonian and canonical transformations}

\paragraph{The quantum LC oscillator}

The LC oscillator of Fig. \ref{lc} can now be treated quantum mechanically. This
circuit with only one active node has a trivial topology. We can immediately
adapt well-known textbook results on the harmonic oscillator. Taking as
variables the integral $\phi $ of the voltage across the inductor and the
corresponding charge $q$ on the capacitor we have the Hamiltonian

\begin{equation}
\mathcal{H}=\frac{q^2}{2C}+\frac{\phi ^2}{2L}  \label{LCHamilt}
\end{equation}

Introducing the usual annihilation and creation operators such that

\begin{equation}
\left[ c,c^{\dagger }\right] =1  \label{annih}
\end{equation}
we have

\begin{eqnarray}
\phi &=&\phi_\mathrm{ZPF} \left( c+c^{\dagger }\right)  \label{ctophi}
\\
q &=&\frac 1i q_\mathrm{ZPF} \left( c-c^{\dagger }\right)
\label{ctoq}
\end{eqnarray}
\begin{equation}
\mathcal{H}=\frac{\hbar \omega _0}2\left( c^{\dagger }c+cc^{\dagger }\right)
=\hbar \omega _0\left( c^{\dagger }c+\tfrac 12\right)  \label{LChamcc}
\end{equation}
where, as in section 1,
\begin{eqnarray}
\omega _0 &=&\sqrt{\frac 1{LC}}  \nonumber \\
Z_0 &=&\sqrt{\frac LC}  \label{OmegandZ}
\end{eqnarray}
and where
\begin{eqnarray}
\phi_\mathrm{ZPF} &=&\sqrt{\frac{\hbar Z_0}{2}}  \nonumber \\
q_\mathrm{ZPF} &=&\sqrt{\frac{\hbar }{2 Z_0}} \label{ZPF}
\end{eqnarray}
represent the standard deviations of the flux and charge fluctuations of the ground state, respectively.
Using Eq.~\ref{Heisen} and the relation

\begin{equation}
\left\langle A\right\rangle =\mathrm{tr}\left[ A\mathrm{e}^{-\beta \mathcal{H%
}}\right] /\mathrm{tr}\left[ \mathrm{e}^{-\beta \mathcal{H}}\right]
\label{Tr}
\end{equation}
where $\beta =\left( k_BT\right) ^{-1}$, we can calculate the flux-flux
correlation function in thermal equilibrium $\left\langle \phi \left(
t\right) \phi \left( 0\right) \right\rangle $. We arrive at

\begin{equation}
\left\langle \phi \left( t\right) \phi \left( 0\right) \right\rangle =\phi_\mathrm{ZPF}^2\left( \left\langle c^{\dagger }c\right\rangle \mathrm{e}%
^{+i\omega _0t}+\left\langle cc^{\dagger }\right\rangle \mathrm{e}^{-i\omega
_0t}\right)  \label{qphi-phi}
\end{equation}
and from

\begin{eqnarray}
\left\langle c^{\dagger }c\right\rangle &=&\frac 1{\mathrm{e}^{\beta \hbar
\omega _0}-1}=\frac 12\coth \tfrac{\beta \hbar \omega _0}2-\frac 12=n\left(
\omega _0\right)  \nonumber \\
\left\langle cc^{\dagger }\right\rangle &=&\frac 1{1-\mathrm{e}^{-\beta
\hbar \omega _0}}=-n\left( -\omega _0\right) =n\left( \omega _0\right) +1
\label{n}
\end{eqnarray}
we get finally

\begin{equation}
\left\langle \phi \left( t\right) \phi \left( 0\right) \right\rangle =\phi_\mathrm{ZPF}^2
\left[ \coth \tfrac{\beta \hbar \omega _0}2\cos \omega _0t-i\sin
\omega _0t\right]  \label{Phicorr}
\end{equation}

Setting $t=0$, we get the variance of flux fluctuations at temperature $T$

\begin{equation}
\left\langle \phi ^2\right\rangle =\frac{\hbar Z_0}2\coth \tfrac{\beta \hbar
\omega _0}2  \label{Qvarphi}
\end{equation}
which interpolates between the zero-point fluctuations result $\left\langle
\phi ^{2}\right\rangle_0 =\phi_\mathrm{ZPF}^2=\hbar Z_{0}/2$ and the high temperature $\left(
k_{B}T\gg \hbar \omega _{0}\right) $ result $\left\langle \phi
^{2}\right\rangle =k_{B}TL$.

From $\left\langle q\left( t\right) q\left( 0\right) \right\rangle
=-C^2d^2\left\langle \phi \left( t\right) \phi \left( 0\right) \right\rangle
/dt^2$ we also get the variance of charge fluctuations

\begin{equation}
\left\langle q^2\right\rangle =\frac \hbar {2Z_0}\coth \tfrac{\beta \hbar
\omega _0}2  \label{Qvarq}
\end{equation}

An important remark can be made: Not only does Eq.~\ref{Phicorr} predict
that the amplitude of fluctuations saturates at low temperature (well-known
zero-point fluctuations) but it also predicts that the quantum correlation
function is not real! The Fourier transform of the correlation function thus
cannot be interpreted as a directly measurable spectral density as it is the
case classically. Let us now discuss the case of a general impedance to
further examine this point.

Introducing the generalized impedance function of an LC oscillator

\begin{eqnarray}
Z_{\mathrm{LC}}\left[ \omega \right] &=&Z_0\left\{ \frac \pi 2\omega _0\left[
\delta (\omega -\omega _0)+\delta (\omega +\omega _0)\right] +\right. 
\nonumber  \label{ZmGen} \\
&&\left. \frac i2\left[ \mathrm{p.p.}\left( \frac{\omega _0}{\omega -\omega
_0}\right) +\mathrm{p.p.}\left( \frac{\omega _0}{\omega +\omega _0}\right) %
\right] \right\}  \label{Zmgen}
\end{eqnarray}
we can rewrite Eq.~\ref{Phicorr} as

\begin{equation}
\left\langle \phi \left( t\right) \phi \left( 0\right) \right\rangle =\frac
\hbar {2\pi }\int \frac{d\omega }\omega \left[ \coth \tfrac{\beta \hbar
\omega }2+1\right] \func{Re}\left( Z_{\mathrm{LC}}\left[ \omega \right]
\right) \exp -i\omega t  \label{qphi-phiLC}
\end{equation}

\subsection{Dissipative quantum circuits}

\subsubsection{The quantum fluctuation-dissipation theorem}

\begin{figure} [htp]
\centering
\includegraphics[angle = 0, width =3.2318in]{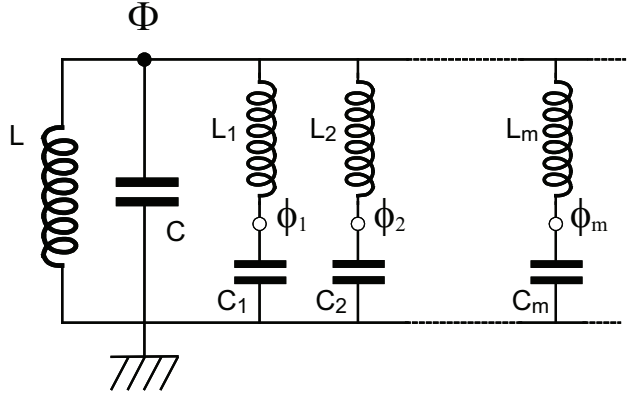}
\caption{\label{lcr-cald-legg}  Caldeira-Leggett representation
of the damped LC circuit of Fig. 1b. }
\end{figure}

We can now obtain the quantum correlation function of the branch flux across
an arbitrary generalized impedance by using the Caldeira-Leggett
representation of Fig.~\ref{lcr-cald-legg}. We simply add the contribution
of all the oscillators and since the correlation function is a linear
function of the real part of the impedance we directly obtain a result of
central importance:

\begin{equation}
\left\langle \Phi \left( t\right) \Phi \left( 0\right) \right\rangle =\frac
\hbar {2\pi }\int_{-\infty }^{+\infty }\frac{d\omega }\omega \left[ \coth 
\tfrac{\beta \hbar \omega }2+1\right] \func{Re}\left( Z\left[ \omega \right]
\right) \exp -i\omega t  \label{Qphi-phi}
\end{equation}

If we now introduce the spectral density of quantum fluctuations

\begin{equation}
S_{\phi \phi}\left[ \omega \right] =\int_{-\infty }^{+\infty }dt\left\langle \Phi \left(
t\right) \Phi \left( 0\right) \right\rangle \exp i\omega t  \label{QS}
\end{equation}
we get the frequency domain relation

\begin{equation}
S_{\phi \phi}\left[ \omega \right] =\frac \hbar \omega \left[ \coth \tfrac{\beta \hbar
\omega }2+1\right] \func{Re}\left( Z\left[ \omega \right] \right)
\label{QFDT}
\end{equation}
which is also called the quantum fluctuation dissipation theorem~\cite{Kubo1966}. Note again that in contrast with a classical spectral density of
fluctuations $S_{\phi \phi}\left[ -\omega \right] \neq S_{\phi \phi}\left[ \omega \right] $. The square brackets have a new meaning here, indicating that both positive and negative frequency arguments have each a separate role, as explained below.

How should we interpret $S_{\phi \phi}\left[ \omega \right] $? To make easier the
comparison with the classical case let us calculate the voltage-voltage
spectral density

\begin{equation}
S_{VV}\left[ \omega \right] =\int_{-\infty }^{+\infty }dt\left\langle \dot{\Phi}%
\left( t\right) \dot{\Phi}\left( 0\right) \right\rangle \exp i\omega t
\label{Svdef}
\end{equation}
which is related to $S_{\phi \phi}\left[ \omega \right] $ by $S_{VV}\left[ \omega \right]
=\omega ^{2}S_{\phi \phi}\left[ \omega \right] $

\begin{equation}
S_{VV}\left[ \omega \right] =\hbar \omega \left[ \coth \tfrac{\beta \hbar \omega }%
2+1\right] \func{Re}\left( Z\left[ \omega \right] \right)  \label{QJohnson}
\end{equation}

In the various limits of interest, $S_{VV}\left[ \omega \right] $ is given by

\begin{equation}
\begin{tabular}{ll}
$\left\vert \hbar \omega \right\vert \ll k_{B}T$ & $S_{VV}\left[ \omega \right]
=2k_{B}T\func{Re}\left( Z\left[ \omega \right] \right) $ \\ 
$\hbar \omega \gg k_{B}T$ & $S_{VV}\left[ \omega \right] =2\hbar \omega \func{Re}%
\left( Z\left[ \omega \right] \right) $ \\ 
$\hbar \omega \ll -k_{B}T$ & $S_{VV}\left[ \omega \right] =0$%
\end{tabular}
\label{limit}
\end{equation}

\paragraph{Interpretation of the quantum spectral density}

The form of $S_{VV}$ in the quantum limit $\left| \hbar \omega \right| \gg k_BT$
shows that the $\omega <0$ part of quantum spectral densities correspond to
processes during which a ``photon'' is transferred from the impedance to the
rest of the circuit while the $\omega >0$ part corresponds to the reverse
process. The quantum fluctuation-dissipation theorem constitutes a
generalization of Planck's black body radiator law. The impedance plays the
role of the black body radiator while the rest of the circuit plays the role
of the atom. Finally, the $\omega <0$ and $\omega >0$ processes correspond to
absorption and emission processes respectively. Note that for $\omega >0$
the $\hbar \omega \func{Re}\left( Z\left[ \omega \right] \right) $ part of $%
S_{VV} $ corresponds to spontaneous emission.

\paragraph{Quantum fluctuations in the damped LC oscillator}

How does dissipation modify the results of Eqs.~\ref{Qvarphi} and~\ref%
{Qvarq}? We can apply the quantum fluctuation-dissipation theorem to
compute the fluctuations of the damped LC oscillator of Fig. 1b. This system
can be represented by the circuit diagram of Fig.~\ref{lcr-cald-legg} in
which we have replaced the admittance $y\left[ \omega \right]$ shunting the main LC oscillator by an
infinite set of series LC oscillators in parallel.

\begin{equation}
\mathcal{H}=\frac{q^2}{2C}+\frac{\phi ^2}{2L}+\sum_m\left[ \frac{q_m^2}{2C_m}%
+\frac{\left( \phi _m-\phi \right) ^2}{2L_m}\right]  \label{LCRhamilt}
\end{equation}

Since this Hamiltonian is quadratic we can in principle find its normal mode
coordinates. However, there is a more efficient method. We can treat the
circuit taken between ground and the closed dot in Fig.~\ref{lcr-cald-legg}
as a dissipative element with an impedance $Z\left[ \omega \right] $ given by

\begin{equation}
Z\left[ \omega \right] =\frac 1{\dfrac 1{jL\omega }+jC\omega +y\left[ \omega
\right]}  \label{Z-y}
\end{equation}

Taking the spanning tree to go through the main inductance $L,$ the node
flux $\phi $ is identical to the flux $\Phi $ through that inductance and we
get

\begin{equation}
\left\langle \Phi ^2\right\rangle =\frac{\hbar Z_0}{2\pi}\int_{-\infty }^{+\infty
}\frac{Z_0\omega _0^2\omega y\left[ \omega \right]}{\left( \omega ^2-\omega
_0^2\right) ^2+Z_0^2\omega _0^2\omega ^2y\left[ \omega \right]^2}\coth 
\tfrac{\beta \hbar \omega }2d\omega  \label{Qvarphiy}
\end{equation}

Similarly, the conjugate charge $q$ is identical to the charge $Q$ on the
main capacitance $C$ and we have

\begin{equation}
\left\langle Q^2\right\rangle =\frac \hbar {2Z_0\pi}\int_{-\infty }^{+\infty }%
\frac{Z_0\omega _0^2\omega ^3y\left[ \omega \right] }{\left( \omega
^2-\omega _0^2\right) ^2+Z_0^2\omega _0^2\omega ^2y\left[ \omega \right] ^2}%
\coth \tfrac{\beta \hbar \omega }2d\omega  \label{Qvarqy}
\end{equation}

We can now apply these results to the so-called Ohmic case (or resistor
case) where the damping admittance is independent of frequency below a
cutoff frequency $\omega _c$ which we take to be much larger than $\omega _0$%
. We take $y\left[ \omega \right]$ of the form

\begin{equation}
y\left[ \omega \right] =\frac 1{R+jL_c\omega }=R^{-1}\frac 1{1-i\dfrac
\omega {\omega _c}}  \label{cutoffdef}
\end{equation}

The integrals in Eqs.~\ref{Qvarphiy} and~\ref{Qvarqy} can be
calculated in closed form~\cite{Grabert1984} and one finds that in the
limit $\omega _c\rightarrow \infty $, $\left\langle \Phi ^2\right\rangle $
becomes independent of $\omega _c$. We have

\begin{equation}
\left\langle \Phi ^2\right\rangle =\hbar Z_0\left\{ \theta +\frac 1{2\pi 
\sqrt{\kappa ^2-1}}\left[ \Psi \left( 1+\lambda _{+}\right) -\Psi \left(
1+\lambda _{-}\right) \right] \right\}  \label{VarLCR}
\end{equation}

where $\Psi \left( x\right) $ is the polygamma function and

\begin{equation}
\theta =\frac{k_BT}{\hbar \omega _0}  \label{thetadef}
\end{equation}

\begin{equation}
\kappa =\left( 2RC\omega _0\right) ^{-1}  \label{kappadef}
\end{equation}

\begin{equation}
\lambda _{\pm }=\frac{\kappa \pm \sqrt{\kappa ^2-1}}{2\pi \theta }
\label{lambdadef}
\end{equation}

In contrast with $\left\langle \Phi ^{2}\right\rangle $, $\left\langle
Q^{2}\right\rangle $ diverges as $\omega _{c}\rightarrow \infty $, a
specifically quantum mechanical result. We have

\begin{equation}
\left\langle Q^2\right\rangle =\frac 1{Z_0^2}\left\langle \Phi
^2\right\rangle +\Delta  \label{VarLCRQ}
\end{equation}

where

\begin{equation}
\Delta =\frac{\hbar \kappa }{\pi Z_0}\left[ 2\Psi \left( 1+\lambda _c\right)
-\frac{\lambda _{+}}{\sqrt{\kappa ^2-1}}\Psi \left( 1+\lambda _{+}\right) +%
\frac{\lambda _{-}}{\sqrt{\kappa ^2-1}}\Psi \left( 1+\lambda _{-}\right) %
\right]  \label{DELTA}
\end{equation}

\begin{equation}
\lambda _c=\frac{\hbar \omega _0}{2\pi k_BT}\left( \frac{\omega _c}{\omega _0%
}-2\kappa \right)  \label{Lambdac}
\end{equation}

\begin{figure} [!th]
\centering
\includegraphics[angle = 0, width =2.4422in]{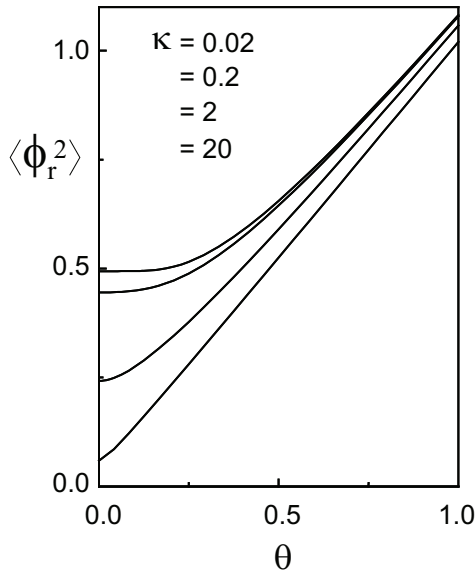}
\caption{\label{phir}  Variations of the dimensionless
variance $\left\langle \protect\phi _{r}^{2}\right\rangle =\left\langle \Phi
^{2}\right\rangle /\left( Z_{0}\hbar \right) $ of flux fluctuations of thespa
LCR circuit as a function of the dimensionless temperature $\protect\theta %
=k_{B}T/\hbar \protect\omega _{0}$ for different values of the dimensionless
damping coefficient $\protect\kappa =\left( 2RC\protect\omega _{0}\right)
^{-1}.$ }
\end{figure}

\begin{figure*} [!ht]
\centering
\includegraphics[angle = 0, width =2.373in]{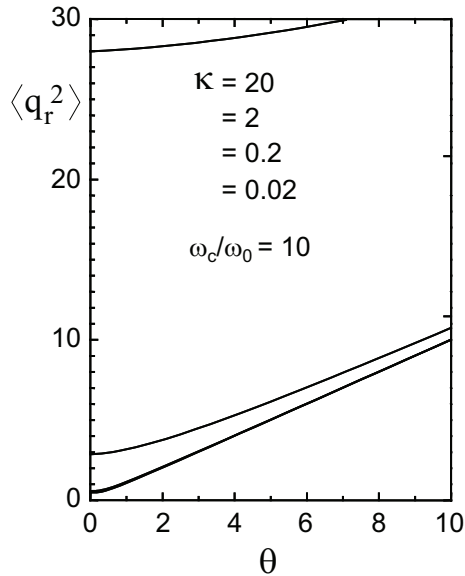}
\caption{\label{qr}  Variations of the dimensionless
variance $\left\langle q_{r}^{2}\right\rangle =Z_{0}\left\langle
Q^{2}\right\rangle /\hbar $ of charge fluctuations of the LCR circuit as a
function of the dimensionless temperature $\protect\theta =k_{B}T/\hbar 
\protect\omega _{0}$ for different values of the dimensionless damping
coefficient $\protect\kappa =\left( 2RC\protect\omega _{0}\right) ^{-1}.$
For all values of $\protect\kappa $, the cutoff frequency $\protect\omega %
_{c}$ of the resistor has been chosen such that $\protect\omega _{c}/\protect%
\omega _{0}=10.$}
\end{figure*}

These expressions are plotted in Figs.~\ref{phir} and~\ref{qr}.

\paragraph{Low temperature limit}

In the limit $\theta \rightarrow 0$ we find the analytical expressions

\begin{equation}
\left\langle \Phi ^2\right\rangle =\frac{\hbar Z_0}2\frac{2\ln \left( \kappa
+\sqrt{\kappa ^2-1}\right) }{\pi \sqrt{\kappa ^2-1}}  \label{LTphi}
\end{equation}

\begin{equation}
\left\langle Q^2\right\rangle =\frac \hbar {2Z_0}\left[ \frac{4\kappa }\pi
\ln \left( \frac{\omega _c}{\omega _0}\right) +\left( 1-2\kappa ^2\right) 
\frac{2\ln \left( \kappa +\sqrt{\kappa ^2-1}\right) }{\pi \sqrt{\kappa ^2-1}}%
\right]  \label{LTQ}
\end{equation}

It is interesting to calculate how the quantum fluctuations depend on the
damping coefficient $\kappa $ in the $\kappa \gg 1$ limit

\begin{eqnarray}
\left\langle \Phi ^2\right\rangle &=&\frac{\hbar Z_0}2\frac{2\ln 2\kappa }{%
\pi \kappa }+\mathcal{O}\left( \frac{\ln \kappa }{\kappa ^3}\right)
\label{philast} \\
\left\langle Q^2\right\rangle &=&\frac \hbar {2Z_0}\frac{4\kappa }\pi \ln
\left( \frac{\omega _c}{2\kappa \omega _0}\right) +\mathcal{O}\left( \frac{%
\ln \kappa }\kappa \right)  \label{qlast}
\end{eqnarray}

We find that the surface of the uncertainty ellipse grows logarithmically
with damping

\begin{equation}
\sqrt{\left\langle \Phi ^2\right\rangle \left\langle Q^2\right\rangle }\sim
\frac \hbar \pi \left[ 2\ln 2\kappa \ln \left( \frac{\omega _c}{2\kappa
\omega _0}\right) \right] ^{\frac 12}  \label{ellipse}
\end{equation}

an effect due to the presence of quantum degrees of freedom inside the
resistor. Apart from that feature, we note that the effect of a resistor on
the quantum mechanical fluctuations of the LC oscillator is essentially to
rescale the size of these fluctuations. We will see in the next section that
the non-linear oscillator formed by a Josephson junction can have a
qualitatively distinct behavior.

\subsubsection{Input-Output theory}

We now present an alternate theory of the damping of a quantum circuit by a dissipative environment. It is based on a role reversal: instead of considering that the circuit loses and gains energy from the environment, we can view the circuit as an elastic scatterer of signals coming from the environment. This approach, nicknamed ``input-output theory'', has the merit of placing the external drive and dissipation of the circuit on the same footing and offers a deeper understanding of the fluctuation dissipation theorem. On the other hand, it simplifies the analysis in the case where the oscillations of the circuit are well within the under-damped regime. This condition makes the environment appear like a resistance (white noise) in the relevant frequency range. This part of the review follows the appendix of Ref.~\cite{Abdo2013} which is itself based on the book by Gardiner and Zoller~\cite{GardinerZoller}.

\begin{figure}[h]
\begin{center}
\includegraphics[width=\columnwidth]{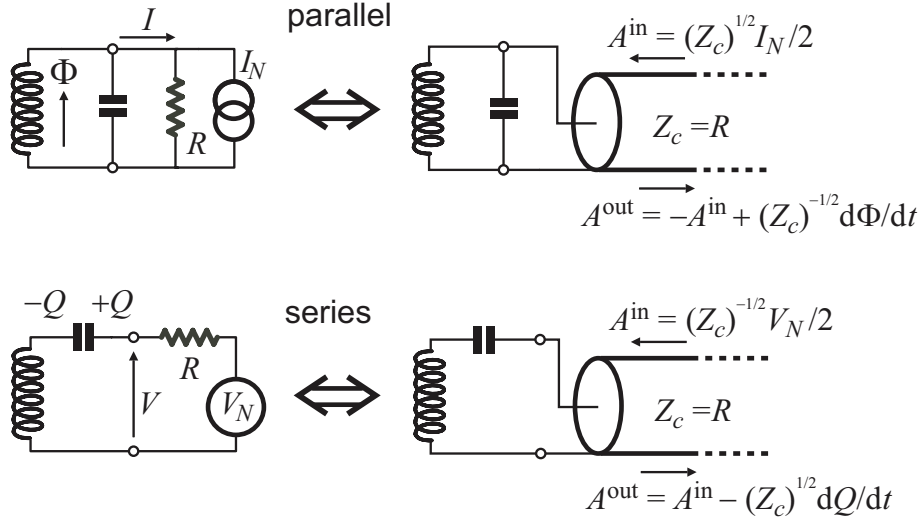}\\
\caption{The damping of a circuit by a resistance $R$ can take place in a
parallel or series way, depending on whether the resistance is placed across a
branch or in series with it. The Nyquist model represents the resistance by a
transmission line with characteristic impedance $Z_{c}=R$. The noise source
associated with the resistance (fluctuation-dissipation theorem) is a parallel
current source in the parallel case and a series voltage source in the series
case. The noise source is replaced in the Nyquist model by incoming thermal
radiation whose amplitude $A^{\mathrm{in}}$ is the square root of the power
flux of the radiation ($A^{\mathrm{in}}$ should not be associated to a vector
potential and is rather like the square root of the length of the Poynting
vector).}%
\label{Nyquist-model}
\end{center}
\end{figure}
%
%
%%%%%%%%%%%%%%%%%%%%%%%%%%%%%%%%%%%%%%%%
\paragraph{Infinite transmission line}
%%%%%%%%%%%%%%%%%%%%%%%%%%%%%%%%%%%%%%%%

Input output theory is based on the Nyquist model of the dissipation by a resistance, in which the environment as replaced by a semi-infinite transmission line (See. Fig.~\ref{Nyquist-model}). Before we treat the coupling between the circuit and this semi-infinite transmission line, let us review the quantization of the field traveling along an infinite transmission line.

The capacitance and inductance per unit length of the line are $C_{\ell}$ and
$L_{\ell}$, respectively. The equations obeyed by the current $I$ along and
the voltage $V$ across the line are
\begin{align}
-\frac{\partial}{\partial x}V\left(  x,t\right)   &  =L_{\ell}\frac{\partial
}{\partial t}I\left(  x,t\right)  ,\label{propagation_eq_1}\\
-\frac{\partial}{\partial x}I\left(  x,t\right)   &  =C_{\ell}\frac{\partial
}{\partial t}V\left(  x,t\right)  , \label{propagation_eq_2}%
\end{align}
in which, for the moment, we treat the fields classically. The characteristic
impedance and propagation velocity are given by
\begin{align}
Z_{c}  &  =\sqrt{\frac{L_{\ell}}{C_{\ell}}},\\
v_{p}  &  =\sqrt{\frac{1}{L_{\ell}C_{\ell}}}.
\end{align}
In order to solve Eqs.~\ref{propagation_eq_1}) and~\ref{propagation_eq_2}),
we introduce two new fields: the left-moving and right-moving wave
amplitudes,
\begin{align}
A^{\rightarrow}\left(  x,t\right)   &  =\frac{1}{2}\left[  \frac{1}%
{\sqrt{Z_{c}}}V\left(  x,t\right)  +\sqrt{Z_{c}}I\left(  x,t\right)  \right]
,\\
A^{\leftarrow}\left(  x,t\right)   &  =\frac{1}{2}\left[  \frac{1}{\sqrt
{Z_{c}}}V\left(  x,t\right)  -\sqrt{Z_{c}}I\left(  x,t\right)  \right]  ,
\end{align}
which have the advantage of treating currents and voltage on the same footing
(note that these amplitudes are not directly related to the vector potential).
The dimension of these fields is [watt]$^{1/2}$ and they are normalized such
that the total power $P$ traversing, in the forward direction, a subsection of
the line at position $x$ and time $t$ is given by
\begin{equation}
P\left(  x,t\right)  =\left[  A^{\rightarrow}\left(  x,t\right)  \right]
^{2}-\left[  A^{\leftarrow}\left(  x,t\right)  \right]  ^{2}. \label{poynting}%
\end{equation}
The quantity $P$ here plays the role of the Poynting vector in full 3D
electrodynamics. Each of the terms at the right hand side of the last equation
is thus the separate contribution of the corresponding wave to the total power flow.

When solving Eqs.~\ref{propagation_eq_1} and~\ref{propagation_eq_2}, we find
\begin{equation}
\frac{\partial}{\partial x}A^{\rightleftarrows}\left(  x,t\right)  =\mp
\frac{1}{v_{p}}\frac{\partial}{\partial t}A^{\rightleftarrows}\left(
x,t\right)  . \label{Eq._of_motion}%
\end{equation}
This relation means that $A^{\rightleftarrows}$ does not depend separately on
$x$ or $t$ but a combination of both and thus:
\begin{align}
A^{\rightarrow}\left(  x,t\right)   &  =A^{\rightarrow}\left(  x=0,t-\frac
{x}{v_{p}}\right)  =A^{\rightarrow}\left(  x-v_{p}t,t=0\right)  ,\nonumber\\
A^{\leftarrow}\left(  x,t\right)   &  =A^{\leftarrow}\left(  x=0,t+\frac
{x}{v_{p}}\right)  =A^{\leftarrow}\left(  x+v_{p}t,t=0\right)  .\nonumber\\
&
\end{align}
The properties of the wave amplitude can be summarized by writing
\begin{align}
A^{\rightleftarrows}\left(  x,t\right)   &  =A_{0}^{\rightleftarrows}\left(
\tau\right)  ,\\
\tau &  =t+\frac{\varepsilon^{\rightleftarrows}}{v_{p}}x,\\
\varepsilon^{\rightleftarrows}  &  =\mp1.
\end{align}
Note that the detailed definition of the retardation $\tau$ depends on the
wave direction. We now turn to the energy density $U\left(  x,t\right)  $,
related to $P$ by the local energy conservation law
\begin{equation}
\frac{\partial U}{\partial t}=-\frac{\partial P}{\partial x}.
\end{equation}
Combining Eqs.~\ref{poynting} and~\ref{Eq._of_motion}, we get
\begin{align}
&  \frac{\partial U\left(  x,t\right)  }{\partial t}\nonumber\\
&  =\frac{2}{v_{p}}\left[  A^{\rightarrow}\left(  x,t\right)  \frac{\partial
}{\partial t}A^{\rightarrow}\left(  x,t\right)  +A^{\leftarrow}\left(
x,t\right)  \frac{\partial}{\partial t}A^{\leftarrow}\left(  x,t\right)
\right]  ,\nonumber\\
&  =\frac{1}{v_{p}}\frac{\partial}{\partial t}\left\{  \left[  A^{\rightarrow
}\left(  x,t\right)  \right]  ^{2}+\left[  A^{\leftarrow}\left(  x,t\right)
\right]  ^{2}\right\}  .
\end{align}
The total energy of the line at time $t$ is, thus,
\begin{equation}
H=\frac{1}{v_{p}}\int_{-\infty}^{+\infty}\left\{  \left[  A^{\rightarrow
}\left(  x,t\right)  \right]  ^{2}+\left[  A^{\leftarrow}\left(  x,t\right)
\right]  ^{2}\right\}  \mathrm{d}x. \label{HinA}%
\end{equation}
When $H$ in Eq.~\ref{HinA} is considered as a functional of dynamical field
variables $A^{\rightarrow}$ and $A^{\leftarrow}$, the equation of motion Eq.~\ref{Eq._of_motion} can be recovered from Hamilton's equation of motion as
\begin{equation}
\frac{\partial}{\partial t}A^{\rightleftarrows}\left(  x,t\right)  =-\left\{
H,A^{\rightleftarrows}\left(  x,t\right)  \right\}  _{P.B.},
\end{equation}
on imposing the Poisson bracket
\begin{align}
&  \left\{  A^{\rightleftarrows}\left(  x_{1},t_{1}\right)
,A^{\rightleftarrows}\left(  x_{2},t_{2}\right)  \right\}  _{P.B.}
  =\frac{1}{2}\frac{\partial}{\partial\left(  \tau_{1}-\tau_{2}\right)
}\delta\left(  \tau_{1}-\tau_{2}\right). \label{PBinA2}%
\end{align}
Therefore, from the classical-quantum correspondence involving the replacement
of Poisson brackets by commutators, we find that the quantum operator version
$\hat{A}^{\rightleftarrows}$ of the fields satisfy the commutation relation
\begin{equation}
\left[  \hat{A}^{\rightleftarrows}\left(  x_{1},t_{1}\right)  ,\hat
{A}^{\rightleftarrows}\left(  x_{2},t_{2}\right)  \right]  =\frac{i\hbar}%
{2}\frac{\partial}{\partial\left(  \tau_{1}-\tau_{2}\right)  }\delta\left(
\tau_{1}-\tau_{2}\right)  ,
\end{equation}
which is analogous to the commutation relation between the electric and
magnetic field in 3-D quantum electrodynamics. Note that the fields are
Hermitian at this stage. Introducing the Fourier transform for signals (whose normalization differs from the previously introduced Fourier transform for response functions),
\begin{equation}
\hat{A}^{\rightleftarrows}\left[  \omega\right]  =\frac{1}{\sqrt{2\pi}}%
\int_{-\infty}^{+\infty}\hat{A}^{\rightleftarrows}\left(  x=0,\tau\right)
e^{i\omega\tau}\mathrm{d}\tau,
\end{equation}
where the Fourier components (which are now non-Hermitian operators) satisfy%

\begin{equation}
\hat{A}^{\rightleftarrows}\left[  \omega\right]  ^{\dag}=A^{\rightleftarrows
}\left[  -\omega\right]  ,
\end{equation}
we can also write the Hamiltonian as
\begin{equation}
\sum_{\sigma=\rightleftarrows}\int_{-\infty}^{+\infty}\hat{A}^{\sigma}\left[
\omega\right]  \hat{A}^{\sigma}\left[  -\omega\right]  \mathrm{d}\omega.
\end{equation}
The field operators in the frequency domain satisfy
\begin{equation}
\left[  \hat{A}^{\rightleftarrows}\left[  \omega_{1}\right]  ,\hat
{A}^{\rightleftarrows}\left[  \omega_{2}\right]  \right]  =\frac{\hbar}%
{4}\left(  \omega_{1}-\omega_{2}\right)  \delta\left(  \omega_{1}+\omega
_{2}\right)  .
\end{equation}
We now introduce the usual quantum field annihilation operators
\begin{align}
a^{\rightarrow}\left[  \omega\right]   &  =\frac{\hat{A}^{\rightarrow}\left[
\omega\right]  }{\sqrt{\hbar\left\vert \omega\right\vert /2}}=a^{\rightarrow
}\left[  -\omega\right]  ^{\dagger},\\
a^{\leftarrow}\left[  \omega\right]   &  =\frac{\hat{A}^{\leftarrow}\left[
\omega\right]  }{\sqrt{\hbar\left\vert \omega\right\vert /2}}=a^{\leftarrow
}\left[  -\omega\right]  ^{\dagger}.
\end{align}
They satisfy the commutation relations
\begin{equation}
\left[  a^{\rightleftarrows}\left[  \omega_{1}\right]  ,a^{\rightleftarrows
}\left[  \omega_{2}\right]  \right]  =\mathrm{sgn}\left(  \frac{\omega
_{1}-\omega_{2}}{2}\right)  \delta\left(  \omega_{1}+\omega_{2}\right)
.\label{bosonic_com}%
\end{equation}
It is useful to note that since
\begin{equation}
a^{\rightleftarrows}\left[  \omega\right]  =a^{\rightleftarrows}\left[
-\omega\right]  ^{\dag},
\end{equation}
Eq.~\ref{bosonic_com} exhaustively describes all possible commutator cases.

In the thermal state of the line, at arbitrary temperature (including $T=0$),

\begin{equation}
\left\langle a^{\rightleftarrows}\left[  \omega_{1}\right]
a^{\rightleftarrows}\left[  \omega_{2}\right]  \right\rangle
=S_{a^{\rightleftarrows}a^{\rightleftarrows}}\left[  \frac{\omega_{1}%
-\omega_{2}}{2}\right]  \delta\left(  \omega_{1}+\omega_{2}\right)  ,
\end{equation}
where%

\begin{equation}
S_{a^{\rightleftarrows}a^{\rightleftarrows}}\left[  \omega\right]
=\mathrm{sgn}\left(  \omega\right)  N_{T}\left(  \omega\right)  .
\end{equation}

These last two equations can be seen as a consequence of the Wiener-Kinchin theorem, and are explained in detail in ref.~\cite{Clerk2010}. 
When $\omega$ is strictly positive $N_{T}\left(  \omega\right)  $ is the
number of available photons per unit bandwidth per unit time traveling on the
line in a given direction around frequency $\omega$%

\begin{align}
N_{T}\left(  \omega\right)   &  =\frac{1}{\exp\left(  \frac{\hbar\omega}%
{k_{B}T}\right)  -1}\\
&  =\frac{1}{2}\left[  \coth\left(  \frac{\hbar\omega}{2k_{B}T}\right)
-1\right]  .
\end{align}
Positive frequencies $\omega$ correspond to the line emitting a photon, while negative frequencies correspond to the line receiving a photon.%

\begin{equation}
N_{T}\left(  -\left\vert \omega\right\vert \right)  =-N_{T}\left(  \left\vert
\omega\right\vert \right)  -1.
\end{equation}
The Bose-Einstein expression $N_{T}\left(  \omega\right)  $ is expected from
the Hamiltonian of the line, which reads, with the $a$ operators,
\begin{equation}
H=\frac{\hbar}{2}\sum_{\sigma=\rightleftarrows}\int_{-\infty}^{+\infty
}\left\vert \omega\right\vert a^{\sigma}\left[  \omega\right]  a^{\sigma
}\left[  -\omega\right]  \mathrm{d}\omega.
\end{equation}
We can now give the expression for the anticommutator of the fields%

\begin{align}
&  \left\langle \left\{  a^{\rightleftarrows}\left[  \omega_{1}\right]
,a^{\rightleftarrows}\left[  \omega_{2}\right]  \right\}  \right\rangle
_{T}=2\mathcal{N}_{T}\left[  \frac{\omega_{1}-\omega_{2}}{2}\right]
\delta\left(  \omega_{1}+\omega_{2}\right) \nonumber\\
&  =\mathrm{sgn}\left(  \frac{\omega_{1}-\omega_{2}}{2}\right)  \coth\left(
\frac{\hbar\left(  \omega_{1}-\omega_{2}\right)  }{4k_{B}T}\right)
\delta\left(  \omega_{1}+\omega_{2}\right)  . \label{bos-anticom6}%
\end{align}

When external drives are present, Eq.~\ref{bos-anticom6} has to be modified with an additional term:
\begin{align}
\mathcal{N}_{T}\left[  \omega\right]   &  =\frac{\mathrm{sgn}\left(
\omega\right)  }{2}\coth\left(  \frac{\hbar\omega}{2k_{B}T}\right)  \\
&  =\mathrm{sgn}\left(  \omega\right)  \left[  N_{T}\left(  \left\vert
\omega\right\vert \right)  +\frac{1}{2}\right]  .
\end{align}
We now introduce the forward-propagating and backward-propagating voltage and
current amplitudes obeying
\begin{align}
V^{\rightarrow}\left(  x,t\right)   &  =\sqrt{Z_{c}}A^{\rightarrow}\left(
x,t\right)  ,\\
V^{\leftarrow}\left(  x,t\right)   &  =\sqrt{Z_{c}}A^{\leftarrow}\left(
x,t\right)  ,
\end{align}%
\begin{align}
I^{\rightarrow}\left(  x,t\right)   &  =V^{\rightarrow}\left(  x,t\right)
/Z_{c},\\
I^{\leftarrow}\left(  x,t\right)   &  =V^{\leftarrow}\left(  x,t\right)
/Z_{c}.
\end{align}
Quantum-mechanically, the voltage and current amplitudes become hermitian
operators
\begin{align}
V^{\rightleftarrows}\left(  x,t\right)   &  \rightarrow\hat{V}%
^{\rightleftarrows}\left(  x,t\right)  ,\\
I^{\rightleftarrows}\left(  x,t\right)   &  \rightarrow\hat{I}%
^{\rightleftarrows}\left(  x,t\right)  .
\end{align}
These operators, in turn, can be expressed in terms of field annihilation
operators as
\begin{align}
\hat{V}^{\rightleftarrows}\left(  x,t\right)   &  =\sqrt{\frac{\hbar Z_{c}%
}{4\pi}}\int_{-\infty}^{+\infty}\mathrm{d}\omega\sqrt{\left\vert
\omega\right\vert }\hat{a}^{\rightleftarrows}\left[  \omega\right]
e^{-i\omega\left(  t\,\mp\,x/v_{p}\right)  },\\
\hat{I}^{\rightleftarrows}\left(  x,t\right)   &  =\sqrt{\frac{\hbar}{4\pi
Z_{c}}}\int_{-\infty}^{+\infty}\mathrm{d}\omega\sqrt{\left\vert \omega
\right\vert }\hat{a}^{\rightleftarrows}\left[  \omega\right]  e^{-i\omega
\left(  t\,\mp\,x/v_{p}\right)  }.
\end{align}
All physical operators can be deduced from these primary expressions. For
instance, the transmission line charge operator, describing the charge in the
line brought from one end to the position $x$, is
\begin{equation}
\hat{Q}^{\rightleftarrows}\left(  x,t\right)  =i\sqrt{\frac{\hbar}{4\pi Z_{c}%
}}\int_{-\infty}^{+\infty}\frac{\mathrm{d}\omega\sqrt{\left\vert
\omega\right\vert }}{\omega}\hat{a}^{\rightleftarrows}\left[  \omega\right]
e^{-i\omega\left(  t\,\mp\,x/v_{p}\right)  }.
\end{equation}
%
%
%%%%%%%%%%%%%%%%%%%%%%%%%%%%%%%%%%%%%%%%%%%%%%%%%%%%%%%%%%%%%%%%%%%%%%%%%%%%%%%%
\paragraph{Nyquist model of resistance: semi-infinite transmission line}
%%%%%%%%%%%%%%%%%%%%%%%%%%%%%%%%%%%%%%%%%%%%%%%%%%%%%%%%%%%%%%%%%%%%%%%%%%%%%%%%
%

We now are in a position to deal with the semi-infinite line extending from
$x=0$ to $x=\infty$, whose terminals at $x=0$ models a resistance $R=Z_{c}$
[see Fig.~\ref{Nyquist-model}]. In that half-line, the left- and right-moving
propagating waves are no longer independent. We will now refer to the wave
amplitude $A^{\leftarrow}\left(  x=0,t\right)  $ as $A^{\mathrm{in}}\left(
t\right)  $ and $A^{\rightarrow}\left(  x=0,t\right)  $ as $A^{\mathrm{out}%
}\left(  t\right)  $. The quantum-mechanical voltage across the terminal of
the resistance and the current flowing into it satisfy the operator relations
\begin{align}
\hat{V}\left(  t\right)   &  =\hat{V}^{\mathrm{out}}\left(  t\right)  +\hat
{V}^{\mathrm{in}}\left(  t\right)  ,\\
\hat{I}\left(  t\right)   &  =\hat{I}^{\mathrm{out}}\left(  t\right)  -\hat
{I}^{\mathrm{in}}\left(  t\right)  .
\end{align}
These relations can be seen either as continuity equations at the interface
between the damped circuit and the resistance/line, or as boundary conditions
linking the semi-infinite line quantum fields $\hat{A}^{\mathrm{in}}\left(
t\right)  $ and $\hat{A}^{\mathrm{out}}\left(  t\right)  $. From the
transmission line relations,
\begin{equation}
\hat{V}^{\mathrm{out},\mathrm{in}}\left(  t\right)  =R\hat{I}^{\mathrm{out}%
,\mathrm{in}}\left(  t\right)  ,
\end{equation}
we obtain
\begin{align}
\hat{I}\left(  t\right)   &  =\frac{1}{R}\hat{V}\left(  t\right)  -2\hat
{I}^{\mathrm{in}}\left(  t\right)  ,\\
&  =\frac{1}{R}\hat{V}\left(  t\right)  -\frac{2}{\sqrt{R}}\hat{A}%
^{\mathrm{in}}\left(  t\right)  .
\end{align}
For a dissipationless circuit with Hamiltonian $H_{bare}\left(  \hat{\Phi
},\hat{Q}\right)  $, where $\hat{\Phi}$ is the generalized flux of the node
electrically connected to the transmission line, and $\hat{Q}$ its canonically
conjugate operator (top panel of Fig.~\ref{Nyquist-model}, we can write the
Langevin equation,
\begin{align}
\frac{\mathrm{d}}{\mathrm{d}t}\hat{Q}  &  =\frac{i}{\hbar}\left[
H_{bare},\hat{Q}\right]  -\hat{I},\nonumber\\
&  =\frac{i}{\hbar}\left[  H_{bare},\hat{Q}\right]  -\frac{\mathrm{d}%
}{R\mathrm{d}t}\hat{\Phi}+\frac{2}{\sqrt{R}}\hat{A}^{\mathrm{in}}\left(
t\right)  . \label{Langevin-example}%
\end{align}
The latter equation is just a particular case of the more general quantum
Langevin equation giving the time evolution of any operator $\hat{Y}$ of a
system with Hamiltonian $H_{bare}$, which is coupled to the semi-infinite
transmission line by an Hamiltonian term proportional to another system
operator $\hat{X}$,
\begin{align}
\frac{\mathrm{d}}{\mathrm{d}t}\hat{Y}  &  =\frac{i}{\hbar}\left[
H_{bare},\hat{Y}\right] \nonumber\\
&  +\frac{1}{2i\hbar}\left\{  \left[  \hat{X},\hat{Y}\right]  ,2R^{\zeta
/2}\hat{A}^{\mathrm{in}}\left(  t\right)  -R^{\zeta}\frac{\mathrm{d}%
}{\mathrm{d}t}\hat{X}\right\}  .\nonumber\\
&  \label{general-QLE1}%
\end{align}
The value of $\zeta$ in Eq.\ref{general-QLE1} depends on whether the
damping is \textquotedblleft parallel" ($\zeta=-1$) or \textquotedblleft
series" type ($\zeta=+1$) [see Fig.~\ref{Nyquist-model}]. In the parallel
case, the greater the line impedance the smaller the damping, whereas in the
series case the situation is reversed.

Equation \ref{general-QLE1} should be supplemented by
\begin{equation}
\left[  \hat{A}^{\mathrm{in}}\left(  t_{1}\right)  ,\hat{A}^{\mathrm{in}%
}\left(  t_{2}\right)  \right]  =\frac{i\hbar}{2}\frac{\partial}%
{\partial\left(  t_{1}-t_{2}\right)  }\delta\left(  t_{1}-t_{2}\right)
\end{equation}
and
\begin{equation}
\hat{A}^{\mathrm{out}}\left(  t\right)  =\zeta\left[  \hat{A}^{\mathrm{in}%
}\left(  t\right)  -R^{\zeta/2}\frac{\mathrm{d}}{\mathrm{d}t}\hat{X}\right]  .
\end{equation}
It follows from the last three equations that the output fields have the same
commutation relation as the input fields
\begin{equation}
\left[  \hat{A}^{\mathrm{out}}\left(  t_{1}\right)  ,\hat{A}^{\mathrm{out}%
}\left(  t_{2}\right)  \right]  =\frac{i\hbar}{2}\frac{\partial}%
{\partial\left(  t_{1}-t_{2}\right)  }\delta\left(  t_{1}-t_{2}\right)  .
\end{equation}
%
%
%%%%%%%%%%%%%%%%%%%%%%%%%%%%%%%%%%%%%%%%%%%%%%%%%%%%%%%%%%%%%%%%%%%%%%%%%%%%%%%%
\paragraph{Quantum Langevin equation in the rotating wave approximation}
%%%%%%%%%%%%%%%%%%%%%%%%%%%%%%%%%%%%%%%%%%%%%%%%%%%%%%%%%%%%%%%%%%%%%%%%%%%%%%%%
%

We now consider an approximate form of the input-output formalism which is
only valid when the system degree of freedom consists of an oscillator with very
low damping, and for which all the frequencies of interest will lie in a
narrow range around the oscillator frequency $\omega_{a}$. We start from Eq.~\ref{Langevin-example} and use
\begin{align}
\hat{\Phi}  &  =\Phi_{ZPF}\left(  a+a^{\dag}\right)  ,\\
\hat{Q}  &  =Q_{ZPF}\frac{\left(  a-a^{\dag}\right)  }{i},
\end{align}
where $\Phi_{ZPF}=\sqrt{\hbar Z_{a}/2}$ and $Q_{ZPF}=\sqrt{\hbar/2Z_{a}}$.

We then obtain, neglecting the effect of driving terms oscillating at twice
the resonance frequency,
\begin{equation}
\frac{\mathrm{d}}{\mathrm{d}t}a=\frac{i}{\hbar}\left[  H_{bare},a\right]
-\omega_{a}\frac{Z_{a}}{2R}a+\sqrt{\frac{2Z_{a}}{\hbar R}}\tilde
{A}^{\mathrm{in}}\left(  t\right)
\end{equation}
with
\begin{equation}
\tilde{A}^{\mathrm{in}}(t)=\int_{0}^{\infty}\hat{A}^{\mathrm{in}}%
[\omega]e^{-i\omega t}\mathrm{d}\omega.
\end{equation}
The field amplitude $\tilde{A}^{\mathrm{in}}(t)$ is non-Hermitian and contains
only the negative frequency component of $A^{\mathrm{in}}(t)$. For signals in
a narrow band of frequencies around the resonance frequency, we can make the
substitution
\begin{equation}
\sqrt{\frac{2}{\hbar\omega_{a}}}\tilde{A}^{\mathrm{in}}\left(  t\right)
\rightarrow\tilde{a}^{\mathrm{in}}\left(  t\right)  ,
\end{equation}
where the frequency components of $\tilde{a}^{\mathrm{in}}(t)$ are equal, in the vicinity of $\omega_{a}$, to those of  
the input field operator $a^{\mathrm{in}}[\omega]$, itself identical to
$a^{\leftarrow}[\omega]$ of the infinite line. We finally arrive at the RWA
quantum Langevin equation, also referred to in the quantum optics literature
as the quantum Langevin equation in the Markov approximation
\begin{equation}
\frac{\mathrm{d}}{\mathrm{d}t}a=\frac{i}{\hbar}\left[  H_{bare},a\right]
-\frac{\gamma_{a}}{2}a+\sqrt{\gamma_{a}}\tilde{a}^{\mathrm{in}}\left(
t\right), \label{QLE}
\end{equation}
where
\begin{equation}
\left[  \tilde{a}^{\mathrm{in}}\left(  t\right)  ,\tilde{a}^{\mathrm{in}%
}\left(  t^{\prime}\right)  ^{\dagger}\right]  =\delta\left(  t-t^{\prime
}\right)  .
\end{equation}
For any oscillator, the input output relationship is obtained from
\begin{equation}
\sqrt{\gamma_{a}}a\left(  t\right)  =\tilde{a}^{\mathrm{in}}\left(  t\right)
-\zeta\tilde{a}^{\mathrm{out}}\left(  t\right)  . \label{IOT}%
\end{equation}
It is worth noting that although $a^{\mathrm{in}}$ and $a^{\mathrm{out}}$ play
the role of $a^{\leftarrow}$ and $a^{\rightarrow}$ in Eq.~\ref{bosonic_com},
only the average values of the moments of $a^{\mathrm{in}}$ can be imposed,
$a^{\mathrm{out}}$ being a \textquotedblleft slave" of the dynamics of
$a^{\mathrm{in}}$, as processed by the oscillator.

\subsection{Measurement operators}

We now introduce the notion that the environment is not completely passive but is able to collect information coming from the system damped by the environment. All quantum measurement experiments on single systems can be analyzed within this framework. Thus, we replace the semi-infinite transmission line of the preceding subsection by a finite transmission terminated by an absorptive detector. This detector performs, of course, measurements on the traveling electromagnetic signal but we can refer its actions to the system itself through Eq.~\ref{IOT}. 

In practice, there are three types of measurement that can be performed: 
\begin{enumerate}
  \item Homodyne measurement, in which the system degree of freedom is analyzed along one component in phase space (i.e. $a+a^\dagger$). Eigenstates of such measurement satisfy a relation of the form $\frac{a+a^\dagger}{2}\ket{I} = I\ket{I}$ where $I$ is the in-phase component of the oscillator, analogous to the position.
  
  \item Heterodyne measurement, in which the system degree of freedom is analyzed along two orthogonal components in phase space (i.e. $a$). In this type of measurement two conjugate operators are measured simultaneously, which necessarily results in added noise. Eigenstates of such measurement satisfy a relation of the form $a\ket{\alpha} = \alpha\ket{\alpha}$ where $\alpha$ is a complex number and $\ket{\alpha}$ is a coherent state. Note that these eigenstates form an over-complete basis, which is another direct result of the commutation relation between two conjugate operators.
  
  \item Photon measurement, in which the system degree of freedom is analyzed in terms of the excitation quanta (i.e. $a^\dagger a$). Eigenstates of such a measurement satisfy a relation of the form $a^\dagger a \ket{n} = n\ket{n}$ where $\ket{n}$ is a Fock state with $n$ photons. 
\end{enumerate}

\subsubsection{The stochastic master equation}

So far, we have discussed the evolution of the quantum circuit under the influence of the Hamiltonian and an external environment interacting with it. We have done this in the operator language, but it is also useful to recast this theory in the language of the density matrix. This leads to the stochastic master equation~\cite{WisemanMilburn,Steck}, which describes the evolution of the density conditioned by the succession of measurement outcomes, also known as the measurement record. An advantage of this formalism is that we end up with an ordinary differential equation in complex number with no non-commuting variables. 

The stochastic master equation can be divided into three parts: the first is the Hamiltonian evolution which is the usual Schrödinger equation, analogous to the Heisenberg equation part of the quantum Langevin equation (Eq.~\ref{QLE}). The second is the Lindblad dissipative evolution, analogous to decay term in the same equation. The last part is the measurement back-action, which corresponds to the stochastic perturbation of the system by a measurement. This term corresponds in the quantum Langevin equation to the influence of the input field. In this formalism, the fluctuation-dissipation theorem is exhibited by the relationship between the back-action of the measurement and the decay of the system.

\newpage
\section{Superconducting Artificial Atoms}

\subsection{The Josephson element}

\subsubsection{The energy operator for the Josephson element}

As we have seen in the introduction, a superconducting tunnel junction can
be modeled by a pure tunnel element (Josephson element) in parallel with a
capacitance. The Josephson element is a pure nonlinear inductance and has an energy operator function of the branch flux $\phi$ given by 
\begin{equation}
h_J(\phi) = E_J \cos\frac{\phi-\phi_\mathrm{offset}}{\phi_0} .
\end{equation}

where $\phi_0 = \hbar/2e$ is the reduced flux quantum and $\phi_\mathrm{offset}$ is an offset branch flux whose role will be discussed later. For the moment suffice to say that its meaning is such that the energy of the element is minimum for $\phi = \phi_\mathrm{offset}$.
In the following we introduce the so-called gauge invariant phase difference $\varphi =  \frac{\phi-\phi_\mathrm{offset}}{\phi_0}$.
The justification of this inductive energy operator is the following:
Consider two isolated superconducting electrodes separated by a thin oxide layer. The electrodes have a number of Cooper pairs $N_1$ and $N_2$ respectively.
While the sum $N_1+N_2$ is conserved, the difference $N = N_1-N_2$ is the degree of freedom of the Cooper-pair tunneling process.     

Quantum-mechanically, $N$ should be treated as an operator $\widehat{N}$
whose eigenstates are macroscopic states of the two electrodes corresponding to
a well-defined number of Cooper pairs having passed through the junction

\begin{equation}
\widehat{N}=\sum_NN\left| N\right\rangle \left\langle N\right|
\label{Nbasis}
\end{equation}

One can show that the tunneling of electrons through the barrier couples the 
$\left| N\right\rangle$ states~\cite{DeGennes1999}. The coupling Hamiltonian is

\begin{equation}
\widehat{h}_{CPT}=-\frac{E_J}2\sum_{N=-\infty }^{+\infty }\left[ \left|
N\right\rangle \left\langle N+1\right| +\left| N+1\right\rangle \left\langle
N\right| \right]  \label{protohamJ}
\end{equation}

The Josephson energy $E_J$ is a macroscopic parameter whose value for BCS
superconductors on both sides of the junction is given by~\cite{Josephson1962} 
\begin{equation}
E_J=\frac 18\frac h{e^2}G_t\Delta  \label{ABJ}
\end{equation}

where $\Delta $ is the superconducting gap and $G_t$ the tunnel conductance
in the normal state. The tunnel conductance is proportional to the
transparency of the barrier and to the surface of the junction.

In the next subsection we are going to show that $\widehat{h}_J$ and $\widehat{h}_{CPT}$ correspond to two representations of the same physical energy.

\subsubsection{The phase difference operator}

Let us now introduce new basis states defined by

\begin{equation}
\left| \theta \right\rangle =\sum_{N=-\infty }^{+\infty }\mathrm{e}%
^{iN\theta }\left| N\right\rangle  \label{deltabasis}
\end{equation}

The index $\theta $ should be thought as the position of a point on the unit
circle since

\begin{equation}
\theta \rightarrow \theta +2\pi  \label{periodic}
\end{equation}

leaves $\left| \theta \right\rangle $ unaffected.

We have conversely 
\begin{equation}
\left| N\right\rangle =\frac 1{2\pi }\int_0^{2\pi }d\theta \ \mathrm{e}%
^{-iN\theta }\left| \theta \right\rangle  \label{delta-N}
\end{equation}

from which we can obtain the expression of $\widehat{h}_{CPT}$ in the $\left|
\theta \right\rangle $ basis

\begin{equation}
\widehat{h}_{CPT}=-\frac{E_J}2\frac 1{2\pi }\int_0^{2\pi }d\theta \left[ \mathrm{%
e}^{i\theta }+\mathrm{e}^{-i\theta }\right] \ \left| \theta \right\rangle
\left\langle \theta \right|  \label{protohamJ1}
\end{equation}

It is natural to introduce the operator

\begin{equation}
\mathrm{e}^{i\widehat{\theta }}=\frac 1{2\pi }\int_0^{2\pi }d\theta \mathrm{e%
}^{i\theta }\ \left| \theta \right\rangle \left\langle \theta \right|
\label{expdel}
\end{equation}

which is such that

\begin{equation}
\mathrm{e}^{i\widehat{\theta }}\left| N\right\rangle =\left| N-1\right\rangle
\label{expdel=trans}
\end{equation}

We can thus write the coupling Hamiltonian (\ref{protohamJ}) as

\begin{equation}
\widehat{h}_{CPT}=-E_J\cos \widehat{\theta }  \label{QJham}
\end{equation}

Thus, if we identify $\varphi \mod 2\pi$ with $\theta$, $\widehat{h}_J$ and $\widehat{h}_{CPT}$ represent the same Hamiltonian.
Note that $\varphi$, as a reduced generalized flux, a pure electromagnetic quantity, takes its values on the whole set of real numbers, whereas $\theta$ is an angle taking its values on the unit circle.
While one might think that $\theta$ and $N$ bears close resemblance to the couple formed by the
number and phase operators for a mode of the electromagnetic field in
quantum optics, it should be stressed that here the pair number
operator takes its eigenvalues in the set of all integers, positive and
negative, whereas the number of photons takes its values in the
set of non-negative integers only. We can write symbolically

\begin{equation}
``\left[ \widehat{\theta},\widehat{N}\right]" =i  \label{commtheta}
\end{equation}

being aware of the fact that due to the compact topology of the manifold $%
\left| \theta \right\rangle$, only periodic functions of $\widehat{\theta }
$ like $\mathrm{e}^{i\widehat{\theta }}$ have a non-ambiguous meaning.

From Eqs.~\ref{der-comq} and~\ref{der-comt} we have

\begin{equation}
\frac d{dt}\widehat{\theta }=\frac 1{i\hbar }\left[ \widehat{\theta },%
\widehat{\mathcal{H}}\right] =-\frac \partial {\hbar \partial \widehat{N}}%
\widehat{\mathcal{H}}  \label{deriv}
\end{equation}

Since $\widehat{N}$ couples linearly to the voltage operator $\widehat{v}$
via the charge $2e$ involved in Cooper-pair transfer, we have

\begin{equation}
\frac d{dt}\widehat{\varphi}=\frac{2e}\hbar \widehat{v}  \label{Jv}
\end{equation}

In the last equation we have used the identity $\frac d{dt}\varphi = \frac d{dt} \theta$.

Using the same type of algebra as in Eqs~\ref{deriv} and~\ref{QJham}, we
find that the current operator $\widehat{i}=2e\frac{d\widehat{N}(t)}{d t}$ is given by

\begin{equation}
\widehat{i}=I_0\sin \widehat{\varphi}  \label{Ji}
\end{equation}

where

\begin{equation}
I_0=\frac{2e}\hbar E_J  \label{I0_EJ}
\end{equation}

Eqs.~\ref{Jv} and~\ref{Ji} together form the quantum constitutive relations
of the Josephson element. 

\subsubsection{Loop combination of several Josephson elements (or a loop formed by a linear inductance and a Josephson element)}

Let us return to the role of the $\phi_\mathrm{offset}$ parameter. One could think that it is a fully inconsequential quantity just like the position of the origin of a coordinate system. 
However, its role appears as soon we have a loop of several inductances, one of which at least being a nonlinear element, such as a Josephson junction. We now introduce the externally imposed
flux $\Phi_\mathrm{ext}$ threading the loop. From Faraday's law, we have $\sum_{b\in \mathrm{loop}} \phi_b = \Phi_\mathrm{ext}$ and thus 
\begin{equation}
\sum_{b \in \mathrm{loop}} \varphi_b = \frac{\Phi_\mathrm{ext}-\sum_b \phi_{b,\mathrm{offset}}}{\phi_0} \label{FluxConstraint}
\end{equation}

Eq.~\ref{FluxConstraint} indicates that around a loop the reduced flux will be in general different from zero, and tunable by the external flux. Thus, the external flux introduces frustration in the system, as not all branches can now have minimal energy. The experimentally observable zero frustration will be obtained when $\Phi_\mathrm{ext} = \Sigma_b  \phi_{b,\mathrm{offset}}$. Therefore, the sum around a loop of the offset fluxes is observable even though each branch value is not. 

Adjusting the frustration in the system by an external flux is crucial in all loop-based quantum circuits, and here we will treat two classic examples. 

\paragraph{Junctions in parallel: The DC SQUID}

The circuit nicknamed ``DC SQUID'' consists of two Josephson tunnel junctions in parallel forming a loop threaded by a flux $\Phi_\mathrm{ext}$. We neglect here the linear inductance of the loop wire. The total inductive energy of this device is 
\begin{equation}
h_{SQUID} = -E_{J_1} \cos \varphi_1 -E_{J_2} \cos \varphi_2
\end{equation}

Where $E_{J_1}$ and $E_{J_2}$ are the respective Josephson energies of the two junctions, and $\varphi_1$ and $\varphi_2$ are the phases across them.
Due to the loop, the phases are related by $\varphi_1 = \Phi_\mathrm{ext}/\phi_0 - \varphi_2$.
Using trigonometric identities, we can recast the equation to a single cosine 

\begin{equation}
h_{SQUID} = -E_{J_\Sigma} \cos(\frac{\Phi_\mathrm{ext}}{2\phi_0})\sqrt{1+d^2 \tan^2 (\frac{\Phi_\mathrm{ext}}{2\phi_0})}\cos \varphi
\end{equation}

Where $E_{J_\Sigma} = E_{J_1} + E_{J_2}$, $d = \frac{E_{J_2}-E_{J_1}}{E_{J_\Sigma}}$ and the new degree of freedom is $\varphi = (\varphi_1+\varphi_2)/2 - \arctan d \tan(\frac{\Phi_\mathrm{ext}}{2\phi_0})$.
Note that this device behaves identically to a single junction, with a tunable Josephson energy. Note also that the offset flux in the definition of $\varphi$ changes with the external flux if the junctions are not identical, but this effect is not directly observable unless the DC SQUID itself is part of a loop.

While the phase potential of the DC SQUID has only even powers of $\varphi$ for any asymmetric and flux, a variation of the asymmetric DC SQUID in which we replace one of the junctions by two larger junctions in series does not have this even symmetry. This circuit is known as the flux qubit (see below). Note that for the right combination of the ratio between the small and large junction and external flux, one can null out the $\varphi^4$ term in the phase potential while maintaining a nonzero $\varphi^3$ term. This property is useful for making a pure three-wave-mixing device.

\paragraph{Junctions in series: Josephson junction arrays}

Let us now consider an array of $M$ identical Josephson junctions in series, each with Josephson energy $E_J$. Let us suppose also that the total reduced $\varphi$ across the array is split equally among the junctions. This hypothesis corresponds to neglecting the effect of the capacitances across the junction, which would allow the current through the Josephson elements to split-off in the array of capacitances.
The total energy can thus be written as
\begin{equation}
h_\mathrm{array}(\varphi) = -M E_J \cos(\varphi/M)
\end{equation}

This equation is strictly valid only when the capacitance array meets two conditions. The capacitance $C_J$ across each junction allows phase slips across it ($\varphi_j \rightarrow \varphi_j + 2\pi$) and therefore must be such that the phase-slip exponential factor~\cite{Matveev2002,Masluk2012} $\exp[-\sqrt{8E_J/E_C}] \ll 1$ where $E_C = e^2/2C_J$ is the Coulomb charging energy of the junction. The other condition stipulates that the capacitance $C_g$ between the array islands and the ground must satisfy $C_J/C_g \gg M$ in order to make the self-resonance of the array above the junction plasma frequency $\sqrt{8E_J E_C}/h$.

In the limit where $M\rightarrow \infty$ this Hamiltonian tends towards
\begin{equation}
h_\mathrm{array}(\varphi) = - \frac{E_J}{2M} \varphi^2 + O(\frac{\varphi^4}{M^3})
\end{equation}

which is the Hamiltonian of a linear inductance whose value is $M L_J$. This allows us to make superinductances, i.e. inductances whose impedances at frequencies below the self resonant frequency are well above the resistance quantum $\hbar/4e^2$, which is impossible with ordinary geometric inductance\cite{Manucharyan2011}.

\subsection{Electromagnetic quantum circuit families}

In this subsection we present a variety of electromagnetic quantum circuits which address various parameter regimes. The different circuits can be distinguished using the two dimension-less ratios $E_J/E_C$ and $(E_J-E_L)/E_L$, where the electrostatic energy $E_C = e^2/2C_\Sigma$ which now includes the total capacitance $C_\Sigma$ shunting the junction, and $E_L = \phi^2_0/L$ is the inductive energy due to an inductance $L$ shunting the junction. 

The first ratio $E_J/E_C$ can be understood as the ``mass'' parameter of the Josephson junction. For $E_J/E_C \ll 1$ the charge having passed through the junction is a good quantum number, and the Cooper-pair tunneling caused by the junction is a small effect. 
For $E_J/E_C \gg 1$ the phase of the junction is a good quantum number and we can expand the phase energy, which we think of as a potential energy, around its minimum value. Thus, the capacitive energy takes the role of the kinetic energy. In this regime one can compute the standard deviation of the phase fluctuations $\sqrt{\left<\varphi^2\right>} = (\frac{2E_C}{E_J})^{1/4}$.

The second ratio $(E_J-E_L)/E_L$ can be understood as being approximately the number of wells minus one in the phase potential, at $\Phi_\mathrm{ext} = \Phi_0/2$. 
The useful circuits are always such that this ratio is positive, which means there are always at least two well at half a flux quantum. This corresponds to the classical potential supporting hysteretic minima.

\begin{figure}[h]
\begin{center}
\includegraphics[width=220 pt]{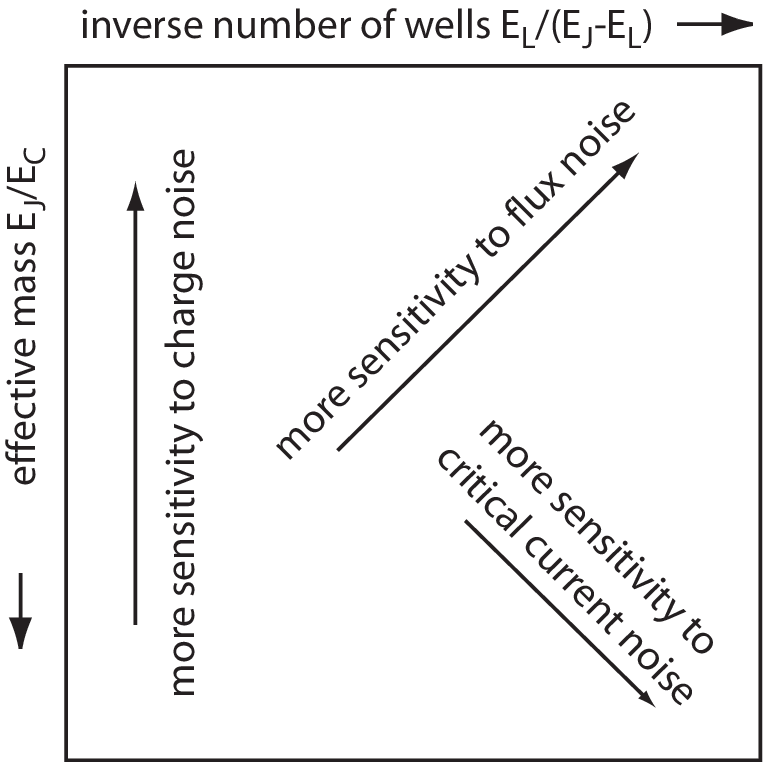}\\
\caption{A sketch of the different mechanisms dominating qubit coherence for different values of $E_J/E_C$ and $E_L/(E_J-E_L)$. In addition to charge noise and flux noise, we are also considering the critical current noise dealt with in Ref.~\cite{DevoretMartinisReview2004}.}%
\label{noisemap}
\end{center}
\end{figure}

In Table~\ref{Juncatalog} we present the different circuits and their place in the circuit ``periodic table'' given by the two ratios described above. We also give a little map (see Fig.~\ref{noisemap}) in parameter space of the different problems plaguing the performance of quantum circuits in the current state-of-the-art. 

\begin{table}[h!]
\centering
\begin{tabular}{  p{1cm} |p{1cm}||p{2.4cm}|p{2cm}|p{2cm}|p{2cm}|}
 \multicolumn{1}{c}{}& \multicolumn{1}{c}{} &\multicolumn{4}{c}{$E_L/(E_J-E_L)$} \\
\cline{3-6}
\multicolumn{1}{c}{} & & $0$ & $\ll$ 1 & $\sim 1$ & $\gg 1$\\
\hhline{~|=====|}
\multirow{4}{*}{$E_J/E_C$} & $\ll 1$ & cooper-pair box  &     &  & \\
\cline{2-6}
& $\sim 1$ & quantronium &   fluxonium  &    &\\
\cline{2-6}
& $\gg 1$ &transmon & & &  flux qubit\\
\cline{2-6}
& $\gg \gg 1$ & & & phase qubit & \\
\cline{2-6}
\end{tabular}
\caption{``periodic table'' of superconducting quantum circuits}
\label{Juncatalog}

\end{table}

\subsubsection{Flux noise and charge noise}
 To understand this map let us write the Hamiltonian for our three element basic circuit including the influence of noise:
 \begin{equation}
 H = 4 E_C(q-q_\mathrm{ext}-q_N(t))^2-E_J \cos(\varphi) + E_L/2  (\varphi-\varphi_\mathrm{ext}-\varphi_N(t))^2 \label{FullJuncHam}
\end{equation}

Here $q$ is the conjugate to $\varphi$ which satisfies $\left[\varphi,q\right] = i$ and $q_N(t)$ and $\varphi_N(t)$ describe the charge and flux noise respectively. Let us point out that $q$, unlike $N$ introduced before, is not the integer number of Cooper pairs having traversed the junction but the charge on the capacitance in units of $2e$. In the last term of Eq.~\ref{FullJuncHam}, the combination $\Phi = \phi_0(\varphi-\varphi_\mathrm{ext}-\varphi_N(t))$ can be understood as the total flux threading loop formed by the inductor and the Josephson element.

Let us first treat the case $E_L \neq 0$, then one can perform a gauge transformation where $q \rightarrow q+q_N(t)$, which leads to the new Hamiltonian:
 \begin{equation}
 H = 4 E_C(q-q_\mathrm{ext})^2-E_J \cos(\varphi) + E_L/2  (\varphi-\varphi_\mathrm{ext}-\varphi_N(t))^2 + \hbar \varphi i_N(t)
\end{equation}

where $i_N(t)$ is the time derivative of $q_N(t)$. Thus we transformed charge noise into offset flux noise. To the usual flux noise influence, we must now add a term related to $i_N(t)$. The expression for the effect of flux and charge fluctuations on qubit coherence can now be expressed as 
\begin{equation}
\frac{1}{T_\varphi} \propto [\frac{\partial \omega_{ge}}{\partial \varphi_\mathrm{ext}}]^2 ((\frac{\hbar \omega}{E_L})^2 S_{qq}[\omega] + S_{\varphi \varphi}[\omega])
\end{equation}

Note that since we are now sensitive to current instead of charge fluctuations, we suppress low frequency charge noise by the factor $\omega^2$~\cite{Koch2009}. This is equivalent to the idea the inductance shunts the charge fluctuations. Notice however, that this suppression is weighted by $E_L^2$ at the denominator, and so as the shunting inductance increases, the effect of charge noise can become dominant.

It is now apparent why $E_L = 0$ is a special case. In this regime we are completely insensitive to flux noise, but we remain sensitive to charge noise through the expression 
\begin{equation}
\frac{1}{T_\varphi} \propto [\frac{\partial \omega_{ge}}{\partial q_\mathrm{ext}}]^2 S_{qq}[\omega] 
\end{equation}

This sensitivity to charge noise can be reduced exponentially by reducing the value of $E_C$, while losing nonlinearity only linearly~\cite{Koch2007}.

\subsubsection{The Cooper pair box}

The Cooper pair box consists simply of a ``small'' Josephson junction ($E_J/E_C \ll 1$) with no shunt inductance, and for which the offset charge $q_\mathrm{ext}$ can be controlled by an external gate voltage. For most of the gate voltage range, the energy eigenstates of this circuit are eigenstates of $q$ since the charging Coulomb energy dominates the Hamiltonian. However, at the special points $q_\mathrm{ext} \mod 1 = 1/2$ the degeneracy between $q=n$ and $q=n+1$ is lifted by the Josephson energy, resulting in a pseudo-spin with Zeeman energy $E_J$~\cite{Bouchiat1998}. The Cooper pair box is the first quantum circuit in which Rabi oscillations between the ground and first excited state have been observed~\cite{Nakamura1999}.

\subsubsection{The transmon}

The transmon~\cite{Koch2007} qubit is a Cooper pair box (with Josephson energy $E_J$ and capacitance $C$) shunted by a large capacitance $C_\mathrm{ext} \gg C$. The capacitances are added into the total capacitance $C_\Sigma = C+C_\mathrm{ext}$, so that the electrostatic energy is significantly reduced ($E_J/E_C \gg 1$). 

The significant benefit of reducing the capacitive energy is removing the sensitivity of the qubit frequency to charge noise. A drift in the charge offset across the junction is screened by the capacitance and no longer changes the transition frequency between the ground and first excited state of the device, thus leading to higher coherence times.

\subsubsection{The flux qubit}

The flux qubit~\cite{Chiorescu2003} is derived from the original proposal by A. J. Leggett of observing macroscopic quantum coherence oscillations between flux states of the RF-SQUID~\cite{Leggett1980, Leggett1987}. Instead of the RF-SQUID, which consists of a Josephson junction shunted by a geometric inductance, the flux qubit consists of a Josephson junction shunted by an effective inductance made up of an array of several bigger Josephson junctions in series (see ``Josephson junction arrays'' above). 

For most external flux bias $\Phi_\mathrm{ext}$ values, the ground state adopted by the system is an eigenstate of the current in the loop and the inductive energy of the circuit (last two terms in Eq.~\ref{FullJuncHam}). At exactly $\Phi_\mathrm{ext} \mod \Phi_0 = \Phi_0/2$ there are two degenerate current states for the device, corresponding to two wells in the potential. This degeneracy is lifted by the Coulomb charging term. The ground-excited state transition energy is a sensitive function of both $\Phi_\mathrm{ext}$ and $E_J$.

The coherence time of the flux qubit is significantly reduced when moving even slightly away from the optimal flux point, due to the high sensitivity of the qubit to flux qubit. To decrease this sensitivity, a variant of the flux qubit~\cite{You2007} has been proposed in which its $E_J/E_C$ ratio is reduced. While decreasing the sensitivity to flux noise, this qubit is now more sensitive to charge noise. To decrease this dependence, a large capacitance is added in parallel with the junction (similar to the transmon qubit), and so this qubit is often called the C-shunt flux qubit~\cite{yan2015}.

\subsubsection{The phase qubit}

The phase qubit~\cite{Martinis1985} is derived from the device on which the first observations of macroscopic quantum energy levels were performed by exploiting the phenomenon of macroscopic quantum tunneling~\cite{Devoret1985}. However, in this qubit the state measurement is performed using flux detection by a DC-SQUID rather than  detection of a DC voltage by a low-noise semiconductor amplifier.
It consists of a large Josephson junction shunted by a geometric inductance, biased to have a metastable potential well. In contrast with other qubits, the Hilbert space for the phase qubit is destroyed by the measurement as the phase particle leaves the metastable well. 

The merit of this qubit is that the signal-to-noise ratio in the readout signal is very high due to the macroscopic tunneling effect.
   
\subsubsection{The fluxonium}

The fluxonium~\cite{Manucharyan2009} artificial atom is a loop circuit made up of a small Josephson junction (with Josephson energy $E_J$ and Coulomb energy $E_C\sim E_J$) in parallel with a large linear inductor, meaning that its inductive energy $E_L$ satisfies $E_L \ll E_J$. The presence of the inductor suppresses the DC component of offset noise as it shunts the two sides of the junction. 

However, a large physical inductance, for example a wire of finite length $L$, is always accompanied by a parasitic capacitance $C_p$. This leads to a $L$-$C_p$ oscillator mode that should not shunt the phase fluctuations of the junction. We thus need to satisfy $(L/C_p)^{1/2} \gg (L_J/C)^{1/2} \sim R_Q$ where $R_Q = \hbar/(2e)^2\approx 1 \mathrm{k\Omega}$ is the resistance quantum. This is impossible to achieve with a geometrical inductance, as its characteristic impedance will always be limited by the vacuum impedance of $377\:\mathrm{\Omega}$. Instead, the fluxonium inductance is implemented using an array of large Josephson junctions (see ``Josephson junction arrays'' above).

The fluxonium level structure strongly depends on the external flux $\Phi_\mathrm{ext}$ across its loop, and this device can be considered a different artificial atom at every flux point. At $\Phi_{ext} = 0$ the low energy fluxonium states are localized inside a single well in flux, and its first excitations are plasma excitation - resembling those of the transmon. At $\Phi_\mathrm{ext} = \Phi_0/2$ the fluxonium low energy states are in two flux wells simultaneously, similar to those of the flux qubit. These are the ``sweet spots'' of the fluxonium, where its energy is first-order insensitive to noise in the external flux and thus the dephasing is minimal. 

\newpage
\section{Conclusions and Perspectives}
The basic concepts of quantum circuits have been discussed in this review. First, a link has been established between standard Quantum Electrodynamics, which deals with how electrons and photons interact in the vacuum, and Josephson circuits in the quantum regime, in which the degrees of freedom are not associated to microscopic particles, but to collective variables of electronic condensed phases of matter which, at low temperatures, can have only a few excitations. The circuit linear inductances and  capacitances form a medium analogous to that of the vacuum supporting the electric and magnetic fields in QED while the role of the Josephson junction corresponds to the non-linear interaction process between electrons and photons. Second, the open system character of quantum circuits has been introduced and we have explained how dissipation of resistances can be dealt with within a quantum-mechanical context. Finally, we have reviewed several examples of key basic circuits and examined the role played by noises in the decoherence of the qubits that can be implemented in these circuits.

Several important topics have been left outside the scope of this review. In particular, the endeavor consisting in using Bloch oscillations at the metrological level has been completely glanced over~\cite{Likharev1985}. In this type of quantum physics, the roles of flux and charge are completely interchanged with respect to their role in commonly used circuits like the transmon. Linked to the question of Bloch oscillation are the proposals of circuits with topological protection from decoherence~\cite{Brooks2013,Doucot2012}. Another fundamental topic of interest, that of driven-dissipative circuits, has also been excluded. In the present review, circuits have been considered as passive devices since no energy was provided to power them. A whole new paradigm is opened when a quantum circuit is submitted to drives at microwave frequencies, which seed the circuit with a bath of photons that can can nourish weak probe signals and dress-up the bare Hamiltonian, giving it entirely new functionalities.  We believe that several of these rich topics will yield important discoveries in the near future, keeping the field of quantum circuits as exciting as when the present review was finished!

\newpage
\section*{Acknowledgements}
\addcontentsline{toc}{section}{Acknowledgements}
The authors are indebted to many colleagues for useful discussions. In particular, we would like to thank Baleegh Abdo, Steven Girvin, Michael Hatridge, Archana Kamal, Angela Kou, Zaki Leghtas, Zlatko Minev, Mazyar Mirrahimi, Ananda Roy, Daniel Sank, Robert Schoelkopf, Shyam Shankar, Clarke Smith, and Evan Zalys-Geller. This work was supported by ARO under Grant No. W911NF-14-1-0011, by MURI-ONR Grant No. N00014-16-1-2270 and by MURI-AFOSR Grand No. FA9550-15-1-0029.

\bibliography{QCreviewbibliography}{}
\bibliographystyle{ieeetr}

\end{document}